\documentclass[journal,final, onecolumn]{IEEEtran}

\usepackage{graphicx}

\usepackage{cite}

\usepackage{color,xcolor}

\ifCLASSINFOpdf

\else

\fi

\usepackage[cmex10]{amsmath}
\usepackage{amssymb}
\usepackage{subeqnarray}
\DeclareMathOperator{\tr}{tr}
\DeclareMathOperator{\cov}{cov}

\allowdisplaybreaks

\begin{document}

\sloppy
%
\title{New Proofs of Extremal Inequalities \\ With Applications}


\author{Yinfei~Xu,~\IEEEmembership{Member,~IEEE,}
        and~Guojun~Chen,~\IEEEmembership{Student~Member,~IEEE}
\thanks{Yinfei Xu and Guojun Chen are with the School of Information Science and Engineering, Southeast University, Nanjing, 210096, China (email:yinfeixu@seu.edu.cn; guojunchen@seu.edu.cn).}
}

\maketitle

\begin{abstract}
The extremal inequality approach plays a key role in network information theory problems. In this paper, we propose a novel monotone path construction in product probability space. The optimality of Gaussian distribution is then established by standard perturbation arguments. The proofs of Liu-Viswanath extremal and vector Generalization of Costa's entropy power inequality are illustrated into the unified framework. As applications, capacity region of the multiple-input multiple-output (MIMO) Gaussian broadcast channel and rate-distortion-equivocation function of the vector Gaussian secure source coding are revisited through our proposed extremal inequality approach.
\end{abstract}


\newtheorem{theorem}{Theorem}
\newtheorem{lemma}{Lemma}
\newtheorem{definition}{Definition}
\newtheorem{remark}{Remark}
\newtheorem{example}{Example}
\newtheorem{corollary}{Corollary}
\newtheorem{proposition}{Propostion}

%

\section{Introduction}
Motivated by multi-antennas communication systems, computing capacity (rate-distortion) region of the vector Gaussian channel (source) is of wide interest. One of the most celebrated results is Weingarten et al.'s solution to capacity of the multiple-input multiple-output
(MIMO) Gaussian broadcast channel \cite{WSS06}. It was  pointed out in \cite{LV07} that Weingarten's result can be represented as optimization problem involving auxiliary random variables, and the capacity problem is reduced to evaluations of the extremal random variables.

Let ${X}, {Z}_{1}, {Z}_{2}$ be independent random vectors taking values in $p$ dimensional real number space $\mathcal{R}^{p}$,
\begin{equation}
\max_{p({x}): \cov ({X}) \preceq \boldsymbol{S}} h({X}+{Z}_{1})- \mu h({X}+{Z}_{2}), \label{eq:00}
\end{equation}
where ${Z}_{1}$ and ${Z}_{2}$ are vector Gaussian random variables with positive definite covariance matrix, and $\mu$ is any real number greater than $1$. In \cite{LV07}, it is shown that the extremal auxiliaries of \eqref{eq:00} are Gaussian distributed. In the proof of extremal inequality, two key techniques are involved. One is the channel enhancement argument \cite{WSS06}, which is exploited to convert ${Z}_{1}$ and ${Z}_{2}$ in degraded order. The other is a strengthened perturbation approach following by Stam \cite{S59} and Blachman \cite{B65} in the proof of entropy power inequality (EPI) \cite{S48}.

The extremal inequality approach is important in its own right. Not only has it paly a key role in characterizing capacity region of the MIMO Gaussian broadcast channel with (without) secrecy \cite{WLSSV09,LLL09,EU12-1,EU12-2,LLPS13,EU13-1,CL14-1,CL14-2,KL14}, but also indispensable to several other vector Gaussian multi-terminal source and channel coding problems \cite{WV07,XCW17,WC13,WC14,EU13,SCT15,XW13,XW16,SCKP10,SP12,SP13,UAZ20, WO11,XC21,XGLC20}.

Note that conventional extremal inequality approach based on enhancement and perturbation might not be flexible enough to include all the situations, \emph{e.g.}, capacity of the MIMO Gaussian broadcast with common and private message. Indeed, Geng and Nair establish the Gaussian optimality through factorization property by regarding the product version of the original channel model \cite{GN14}. Another example is on rate-distortion-equivocation function evaluation of the vector Gaussian secure source coding. The conventional approach can had resisted solution beyond the case of $\mu=1$ \cite{EU13}.

Inspired by the recent work of \cite{WC19,XGLC20}, we construct the Gaussian perturbation variable via a continuously parameterized tensorization process. Although the proposed method works in the product probability space as in \cite{GN14}, it is monotone path centric, and enables to leverage standard perturbation techniques  \cite{LV07,XW13} to prove the optimality of the Gaussian solution. To show flexibility and informatics of our proposed construction, we will recover some existing extremal inequalities under the monotone path arguments, including Liu-Viswanath extremal inequality \cite{LV07} and vector generalization of Costa's EPI \cite{LLPS13}.

The rest of paper is organized as follows. In Section \ref{sec2}, we represent our techniques on Liu-Viswanath extremal inequality, and show how to prove Gaussian optimality without channel enhancement argument. In Section \ref{sec3}, we illustrate the similar idea can also recover the vector generalization of Costa's EPI. We further revisit the main result of \cite{GN14} and \cite{EU13} in Section \ref{sec4} and \ref{sec5}, separately.
The fundamental limits of the two mulit-user information theory problems are fully characterized via our perturbation methods. Finally, we conclude by summarizing our contributions in the context of applications in mulit-user information theory problems.

\section{Liu-Viswanath Extremal Inequality} \label{sec2}

In this section, we provide an alternative proof of Liu-Viswanath extremal inequality in details.
\begin{theorem}\cite[Theorem 1]{LV07}\label{thm1}
Let $Z_{1}$, $Z_{2}$ be Gaussian random vectors with positive semi-definite covariance matrices $\boldsymbol{K}_{1}$ and $\boldsymbol{K}_{2}$, respectively. If there exists a positive semi-definite matrix $\boldsymbol{B}^{*}$ such that
\begin{align}
\left(   \boldsymbol{B}^{*} + \boldsymbol{K}_{1}   \right)^{-1} + \boldsymbol{M}_{1} = \mu \left(   \boldsymbol{B}^{*} + \boldsymbol{K}_{2}   \right)^{-1} + \boldsymbol{M}_{2},
\end{align}
for $\mu \geq 1$, and $\boldsymbol{M}_1$, $\boldsymbol{M}_2$, $\boldsymbol{S}$ are positive semi-definite matrices satisfying
\begin{align}
\boldsymbol{B}^{*}\boldsymbol{M}_{1} = 0,\\
\left(\boldsymbol{S} - \boldsymbol{B}^{*} \right) \boldsymbol{M}_2 =0,
\end{align}
we have
\begin{equation}
h(X+Z_1)-\mu h(X+Z_2) \leq \frac{1}{2} \log \left|  (2 \pi e) \left(\boldsymbol{B}^{*} + \boldsymbol{K}_{1}\right)          \right| - \frac{\mu}{2} \log \left|  (2 \pi e)\left( \boldsymbol{B}^{*} + \boldsymbol{K}_{2}   \right)       \right|,\label{eq:main}
\end{equation}
for any $X$ such that $\cov (X) \preceq \boldsymbol{S}$.
\end{theorem}
\begin{remark}
In \cite{LV07}, the extremal inequality was proved in two steps: Firstly, Gaussian random vector $Z_{1}$ is transformed into $\tilde{Z}_{1}$ via enhancement argument \cite{WLSSV09}, which is degraded to $Z_{2}$. Secondly, a monotone path centered approach is invoked to show that Gaussian optimality of the auxiliary optimization problem on arbitrary distributed random vectors. Alternatively, Geng and Nair introduced a doubling trick to obtain Liu-Viswanath extremal inequality directly in \cite{GN14}. Motivated by the doubling trick, we construct an appropriate monotone path in the tensorized probability space, and show the Gaussian optimality without enhancement argument.
\end{remark}

\subsection{Monotone Path Construction}

We consider the {covariance preserved transform} in \cite{DCT91}. Specifically, for any $\gamma \in (0,1)$, define
\begin{align}
X_{+,\gamma} =\sqrt{1-\gamma}X + \sqrt{\gamma}X^{G}, \label{eq:trans1}\\
X_{-,\gamma} =\sqrt{\gamma}X - \sqrt{1-\gamma}X^{G}, \label{eq:trans2}
\end{align}
where $X^{G}$ follows Gaussian distribution $\mathcal{N}(0, \boldsymbol{B}^{*})$, which is independent of $X$. Notice that $\{X_{+,\gamma}, X_{-,\gamma}\}$ is a family of distributions connecting arbitrary distributed $X$ to Gaussian distributed $X^{G}$. Let $g(\gamma)$ be the objective function evaluated along the perturbed path $\{X_{+,\gamma}, X_{-,\gamma}\}$,
\begin{align}
g(\gamma) =&  \mu h\left(X_{+,\gamma}+\sqrt{1-\gamma}N_{1} + \sqrt{\gamma} N^{G}_{1},X_{-,\gamma}+ \sqrt{\gamma}N_{2}
- \sqrt{1-\gamma}N^{G}_{2}\right) \nonumber \\
&-(\mu-1) h \left(  X_{+,\gamma}+\sqrt{1-\gamma}N_{1} + \sqrt{\gamma} N^{G}_{1}        \right). \label{eq:g_gamma}
\end{align}
In \eqref{eq:g_gamma}, $N_{1}^{G}$, $N_{2}^{G}$ are Gaussian random vectors with the same distribution of $N_1$, $N_2$, which are independent of $N_1$, $N_2$.

If $\gamma =0$, we have
\begin{align}
g(0) = & \mu h(X + {N}_{1}, -X^{G} - {N}^{G}_{2})- (\mu-1) h(X+{N}_{1})  \\
     = & h(X+{N}_{1}) + \frac{\mu}{2} \log \left|  (2 \pi e)\left( \boldsymbol{B}^{*} + \boldsymbol{K}_{2}   \right)       \right|.
\end{align}
If $\gamma =1$, we have
\begin{align}
g(1) = & \mu h(X^{G}+N^{G}_{1},X+N_2)-(\mu-1)h(X^{G} + N^{G}_{1}) \\
     = & \mu h(X+N_{2}) + \frac{1}{2}\log \left|  (2 \pi e)\left( \boldsymbol{B}^{*} + \boldsymbol{K}_{1}   \right)       \right|.
\end{align}

\subsection{Derivative Evaluation}

For the sake of simplifying notations, we denote by
\begin{align}
S_{1} &\triangleq \sqrt{\gamma}X^{G}+\sqrt{1-\gamma}N_{1} + \sqrt{\gamma} N^{G}_{1}, \label{eq:S1}\\
S_{2} &\triangleq -\sqrt{1-\gamma}X^{G}+ \sqrt{\gamma}N_{2}- \sqrt{1-\gamma}N^{G}_{2}. \label{eq:S2}
\end{align}
The covariance matrices of $S_{1}$ and $S_{2}$ are
\begin{align}
\boldsymbol{K}_{S_1} &= \gamma \boldsymbol{B}^{*} + \boldsymbol{K}_{1},\\
\boldsymbol{K}_{S_{2}} &= (1-\gamma) \boldsymbol{B}^{*} + \boldsymbol{K}_{2}.
\end{align}

It is easily verified that
\begin{align}
\sqrt{\gamma}S_{1} - \sqrt{1-\gamma}S_{2} &= X^{G} + \sqrt{\gamma(1-\gamma)} {N}_{1} + {\gamma}N^{G}_{1} - \sqrt{\gamma(1-\gamma)} {N}_{2}+ (1-\gamma)N^{G}_{2}.
\end{align}
The covariance of $\sqrt{\gamma}S_{1} - \sqrt{1-\gamma}S_{2}$ is
\begin{align}
\boldsymbol{K}_{\Delta} &=\boldsymbol{B}^{*} + \gamma \boldsymbol{K}_{1} + (1-\gamma) \boldsymbol{K}_{2} \label{eqn:KW}\\
&=\boldsymbol{K}_{S_1} + \boldsymbol{K}_{S_2}.
\end{align}

We can thereby represent $\sqrt{\gamma}S_1$ by using its optimal estimation when giving $\sqrt{\gamma}S_{1} - \sqrt{1-\gamma}S_{2}$, as below,
\begin{align}
\sqrt{\gamma} S_1 = \gamma (\boldsymbol{B}^{*} + \boldsymbol{K}_{1}) \boldsymbol{K}_{\Delta}^{-1}\left(    \sqrt{\gamma}S_{1} - \sqrt{1-\gamma}S_{2}  \right) + \sqrt{\gamma (1-\gamma)} W ,\label{eq:W}
\end{align}
where $W$ is independent of $\sqrt{\gamma}S_{1} - \sqrt{1-\gamma}S_{2}$, with covariance
\begin{align}
\boldsymbol{K}_{W} &=\boldsymbol{K}_{S_1} - \gamma \left( \boldsymbol{B}^{*} +\boldsymbol{K}_{1}\right)\boldsymbol{K}_{\Delta}^{-1}\left( \boldsymbol{B}^{*} +\boldsymbol{K}_{1}\right) \\
  &=\left(   {(1-\gamma)}(\boldsymbol{B}^{*} + \boldsymbol{K}_{1})^{-1} +  \gamma (\boldsymbol{B}^{*} + \boldsymbol{K}_{2})^{-1}           \right)^{-1} - \boldsymbol{B}^{*}\\
  &=(\boldsymbol{B}^{*} + \boldsymbol{K}_{2}) \boldsymbol{K}^{-1}_{\Delta}(\boldsymbol{B}^{*} + \boldsymbol{K}_{1}) - \boldsymbol{B}^{*}. \label{eq:KW3}
\end{align}

Calculating the derivative of the first term of bivariate differential entropy in \eqref{eq:main} on $\gamma$, it can be written as
\begin{align}
&\frac{d}{d\gamma}h\left(X_{+,\gamma}+\sqrt{1-\gamma}N_{1} + \sqrt{\gamma} N^{G}_{1},X_{-,\gamma}+ \sqrt{\gamma}N_{2}- \sqrt{1-\gamma}N^{G}_{2}\right)\nonumber \\
&\overset{(a)}=\frac{d}{d\gamma} h \left(   \sqrt{1-\gamma}X+S_{1}, \sqrt{\gamma}X+S_2         \right) \label{eq:ddev1}\\
&= \frac{d}{d\gamma}\left\{ h \left(   \sqrt{\gamma(1-\gamma)}X+\sqrt{\gamma}S_{1}, \sqrt{\gamma}S_{1} - \sqrt{1-\gamma} S_{2}       \right)+\frac{n}{2} \log \left(\gamma(1-\gamma)\right) \right\} \\
&=\frac{d}{d\gamma}\left\{ h \left( \left.  \sqrt{\gamma(1-\gamma)}X+\sqrt{\gamma}S_{1}\right| \sqrt{\gamma}S_{1} - \sqrt{1-\gamma} S_{2}       \right)+  h\left(\sqrt{\gamma}S_{1} - \sqrt{1-\gamma} S_{2}\right)+     \frac{n}{2} \log \left(\gamma(1-\gamma)\right) \right\} \\
&\overset{(b)}=\frac{d}{d\gamma}\left\{  h(X+W)    +\frac{1}{2} \log \left| (2 \pi e)   \boldsymbol{K}_{\Delta}     \right|                                            \right\} \\
& \overset{(c)}=\frac{1}{2}\tr \left\{      \left(   \nabla_\gamma \boldsymbol{K}_{W}  \right)    J(X+W)           \right\} +\frac{1}{2} \tr \left\{ \left(\boldsymbol{K}_{1}  - \boldsymbol{K}_{2}\right)  \boldsymbol{K}^{-1}_{\Delta}       \right\} \\
& \overset{(d)}=\frac{1}{2(1-\gamma)^2}\tr \left\{  \left(     (\boldsymbol{B}^{*} + \boldsymbol{K}_{1})^{-1} - (\boldsymbol{B}^{*} + \boldsymbol{K}_{2})^{-1}  \right) \left(   (1-\gamma)^{2}    (\boldsymbol{B}^{*} + \boldsymbol{K}_{1}) \boldsymbol{K}^{-1}_{\Delta}(\boldsymbol{B}^{*} + \boldsymbol{K}_{2}) J(X+W) \right.\right.\nonumber \\
& \qquad \qquad \qquad \qquad \left.\left.(\boldsymbol{B}^{*} + \boldsymbol{K}_{2}) \boldsymbol{K}^{-1}_{\Delta}(\boldsymbol{B}^{*} + \boldsymbol{K}_{1})  -(1-\gamma)^{2}(\boldsymbol{B}^{*} + \boldsymbol{K}_{1}) \boldsymbol{K}^{-1}_{\Delta}(\boldsymbol{B}^{*} + \boldsymbol{K}_{2})     \right)         \right\}\label{eq:dev1}
\end{align}
where
\begin{enumerate}
\item [(a)] is due to notations of $S_1$ and $S_2$ in \eqref{eq:S1} and \eqref{eq:S2};
\item[(b)] is due to \eqref{eqn:KW} and \eqref{eq:W};
\item[(c)] is due to chain rule of matrix calculus and Lemma \ref{de} in Appendix \ref{app_lea2};
\item[(d)] is due to following calculations:
  \begin{align}
   \nabla_\gamma \boldsymbol{K}_{W}  &= \left(   {(1-\gamma)}(\boldsymbol{B}^{*} + \boldsymbol{K}_{1})^{-1} +  \gamma (\boldsymbol{B}^{*} + \boldsymbol{K}_{2})^{-1}           \right)^{-1}\left(     (\boldsymbol{B}^{*} + \boldsymbol{K}_{1})^{-1} - (\boldsymbol{B}^{*} + \boldsymbol{K}_{2})^{-1}            \right) \nonumber \\
   &\quad \left(   {(1-\gamma)}(\boldsymbol{B}^{*} + \boldsymbol{K}_{1})^{-1} +  \gamma (\boldsymbol{B}^{*} + \boldsymbol{K}_{2})^{-1}           \right)^{-1} \nonumber \\
   &= (\boldsymbol{B}^{*} + \boldsymbol{K}_{2}) \boldsymbol{K}^{-1}_{\Delta}(\boldsymbol{B}^{*} + \boldsymbol{K}_{1})\left(     (\boldsymbol{B}^{*} + \boldsymbol{K}_{1})^{-1} - (\boldsymbol{B}^{*} + \boldsymbol{K}_{2})^{-1}  \right)(\boldsymbol{B}^{*} + \boldsymbol{K}_{1}) \boldsymbol{K}^{-1}_{\Delta}(\boldsymbol{B}^{*} + \boldsymbol{K}_{2}), \\
  \boldsymbol{K}_{1} - \boldsymbol{K}_{2}&= (\boldsymbol{B}^{*} + \boldsymbol{K}_{1}) - (\boldsymbol{B}^{*} + \boldsymbol{K}_{2})\nonumber \\
  &=-(\boldsymbol{B}^{*} + \boldsymbol{K}_{2})\left(     (\boldsymbol{B}^{*} + \boldsymbol{K}_{1})^{-1} - (\boldsymbol{B}^{*} + \boldsymbol{K}_{2})^{-1}  \right)(\boldsymbol{B}^{*} + \boldsymbol{K}_{1}).\label{eq:dev2}
  \end{align}
\end{enumerate}
Calculating the derivative of the second term of differential entropy in \eqref{eq:main} on $\gamma$, it can be written as
\begin{align}
&\frac{d}{d\gamma}h\left(X_{+,\gamma}+\sqrt{1-\gamma}N_{1} + \sqrt{\gamma} N^{G}_{1}\right)\nonumber \\
&\overset{(a)}=\frac{d}{d\gamma} h\left(    \sqrt{1-\gamma}X + S_{1}               \right) \label{eq:ddev2}\\
&=\frac{d}{d\gamma} \left\{  h\left(  X + \frac{1}{\sqrt{1-\gamma}}S_{1}  \right)+\frac{n}{2}\log (1-\gamma)  \right\}  \\
&\overset{(b)} = \frac{1}{2(1-\gamma)^2} \tr \left\{     (\boldsymbol{B}^{*} + \boldsymbol{K}_{1}) J\left( X + \frac{1}{\sqrt{1-\gamma}}S_{1}\right)         \right\} -\frac{n}{2(1-\gamma)} \\
&=\frac{1}{2(1-\gamma)^{2}} \tr \left\{     (\boldsymbol{B}^{*}+\boldsymbol{K}_{1})^{-1} \left( {(\boldsymbol{B}^{*}+\boldsymbol{K}_{1})}        J\left(X+\frac{1}{\sqrt{1-\gamma}}S_{1}\right)  {(\boldsymbol{B}^{*}+\boldsymbol{K}_{1})} - {(1-\gamma)} (\boldsymbol{B}^{*}+\boldsymbol{K}_{1})                    \right)    \right\}\label{eq:ddevv2}
\end{align}
where
\begin{enumerate}
\item [(a)] is due to notations of $S_1$ in \eqref{eq:S1};
\item [(b)] is due to Lemma \ref{de} in Appendix \ref{app_lea2} and following calculation:
\begin{align}
\nabla_\gamma \left(   \frac{\boldsymbol{K}_{S_1}}{1-\gamma}  \right)  &= \nabla_\gamma \left(   \frac{\gamma \boldsymbol{B}^{*}+\boldsymbol{K}_{1}}{1-\gamma}  \right) = \frac{\boldsymbol{B}^{*}+\boldsymbol{K}_{1}}{(1-\gamma)^{2}}.
\end{align}
\end{enumerate}

Combining \eqref{eq:g_gamma}, \eqref{eq:dev1} and \eqref{eq:dev2}, we obtain
\begin{align}
& (1-\gamma)^2 \frac{d}{d\gamma}g(\gamma)\nonumber \\
& = \tr \left\{  \left(     \mu (\boldsymbol{B}^{*} + \boldsymbol{K}_{1})^{-1} - \mu (\boldsymbol{B}^{*} + \boldsymbol{K}_{2})^{-1}  \right) \left(   (1-\gamma)^{2}    (\boldsymbol{B}^{*} + \boldsymbol{K}_{1}) \boldsymbol{K}^{-1}_{\Delta}(\boldsymbol{B}^{*} + \boldsymbol{K}_{2}) J(X+W) \right.\right.\nonumber \\
& \qquad \qquad \qquad \qquad \left.\left.(\boldsymbol{B}^{*} + \boldsymbol{K}_{2}) \boldsymbol{K}^{-1}_{\Delta}(\boldsymbol{B}^{*} + \boldsymbol{K}_{1})  -(1-\gamma)^{2}(\boldsymbol{B}^{*} + \boldsymbol{K}_{1}) \boldsymbol{K}^{-1}_{\Delta}(\boldsymbol{B}^{*} + \boldsymbol{K}_{2})     \right)         \right\}\nonumber\\
& \quad - \tr \left\{   (\mu-1)  (\boldsymbol{B}^{*}+\boldsymbol{K}_{1})^{-1} \left( {(\boldsymbol{B}^{*}+\boldsymbol{K}_{1})}        J\left(X+\frac{1}{\sqrt{1-\gamma}}S_{1}\right)  {(\boldsymbol{B}^{*}+\boldsymbol{K}_{1})} - {(1-\gamma)} (\boldsymbol{B}^{*}+\boldsymbol{K}_{1})                    \right)    \right\}. \label{eq:low}
\end{align}

\subsection{Lower Bounds}
In this subsection, we will show that \eqref{eq:low} is lower bounded by $0$. By \eqref{eq:W}, we firstly notice that
\begin{equation}
\frac{1}{\sqrt{1-\gamma}} S_1 = W+\sqrt{\frac{\gamma}{1-\gamma}} (\boldsymbol{B}^{*} + \boldsymbol{K}_{1}) \boldsymbol{K}_{\Delta}^{-1}\left(    \sqrt{\gamma}S_{1} - \sqrt{1-\gamma}S_{2}  \right).
\end{equation}
Applying Fisher information inequality of Lemma \ref{fi_inq} in Appendix \ref{app_lea2}, we obtain
\begin{align}
&{(\boldsymbol{B}^{*}+\boldsymbol{K}_{1})}\boldsymbol{K}^{-1}_{\Delta} \boldsymbol{K}_{\Delta}       J\left(X+\frac{1}{\sqrt{1-\gamma}}S_{1}\right) \boldsymbol{K}_{\Delta}\boldsymbol{K}^{-1}_{\Delta} {(\boldsymbol{B}^{*}+\boldsymbol{K}_{1})} - {(1-\gamma)} (\boldsymbol{B}^{*}+\boldsymbol{K}_{1})            \nonumber \\
& \preceq      (\boldsymbol{B}^{*} + \boldsymbol{K}_{1}) \boldsymbol{K}^{-1}_{\Delta}(1-\gamma)(\boldsymbol{B}^{*} + \boldsymbol{K}_{2}) J(X+W)(1-\gamma)(\boldsymbol{B}^{*} + \boldsymbol{K}_{2}) \boldsymbol{K}^{-1}_{\Delta}(\boldsymbol{B}^{*} + \boldsymbol{K}_{1}) \nonumber \\
& \quad +\gamma (1-\gamma)(\boldsymbol{B}^{*} + \boldsymbol{K}_{1}) \boldsymbol{K}^{-1}_{\Delta} (\boldsymbol{B}^{*} + \boldsymbol{K}_{1})- {(1-\gamma)} (\boldsymbol{B}^{*}+\boldsymbol{K}_{1})\\
& =   (1-\gamma)^{2}    (\boldsymbol{B}^{*} + \boldsymbol{K}_{1}) \boldsymbol{K}^{-1}_{\Delta}(\boldsymbol{B}^{*} + \boldsymbol{K}_{2}) J(X+W)(\boldsymbol{B}^{*} + \boldsymbol{K}_{2}) \boldsymbol{K}^{-1}_{\Delta}(\boldsymbol{B}^{*} + \boldsymbol{K}_{1})\nonumber \\ &\quad-(1-\gamma)^{2}(\boldsymbol{B}^{*} + \boldsymbol{K}_{1}) \boldsymbol{K}^{-1}_{\Delta}(\boldsymbol{B}^{*} + \boldsymbol{K}_{2}).
\end{align}
Thus, \eqref{eq:low} can be lower bounded by
\begin{align}
\frac{d}{d\gamma}g(\gamma)\geq & \tr \left\{  \left(      (\boldsymbol{B}^{*} + \boldsymbol{K}_{1})^{-1} - \mu (\boldsymbol{B}^{*} + \boldsymbol{K}_{2})^{-1}  \right) \left(      (\boldsymbol{B}^{*} + \boldsymbol{K}_{1}) \boldsymbol{K}^{-1}_{\Delta}(\boldsymbol{B}^{*} + \boldsymbol{K}_{2}) J(X+W) \right.\right.\nonumber \\
& \qquad \left.\left.(\boldsymbol{B}^{*} + \boldsymbol{K}_{2}) \boldsymbol{K}^{-1}_{\Delta}(\boldsymbol{B}^{*} + \boldsymbol{K}_{1})  -(\boldsymbol{B}^{*} + \boldsymbol{K}_{1}) \boldsymbol{K}^{-1}_{\Delta}(\boldsymbol{B}^{*} + \boldsymbol{K}_{2})     \right)         \right\}\\
= & \tr \left\{  \boldsymbol{M}_{2} \left(      (\boldsymbol{B}^{*} + \boldsymbol{K}_{1}) \boldsymbol{K}^{-1}_{\Delta}(\boldsymbol{B}^{*} + \boldsymbol{K}_{2}) J(X+W)(\boldsymbol{B}^{*} + \boldsymbol{K}_{2}) \boldsymbol{K}^{-1}_{\Delta}(\boldsymbol{B}^{*} + \boldsymbol{K}_{1}) \right.\right.\nonumber \\
& \qquad \left.\left.  -(\boldsymbol{B}^{*} + \boldsymbol{K}_{1}) \boldsymbol{K}^{-1}_{\Delta}(\boldsymbol{B}^{*} + \boldsymbol{K}_{2})     \right)         \right\} \label{eq:M_2}\\
& - \tr \left\{  \boldsymbol{M}_{1}\left(      (\boldsymbol{B}^{*} + \boldsymbol{K}_{1}) \boldsymbol{K}^{-1}_{\Delta}(\boldsymbol{B}^{*} + \boldsymbol{K}_{2}) J(X+W) (\boldsymbol{B}^{*} + \boldsymbol{K}_{2}) \boldsymbol{K}^{-1}_{\Delta}(\boldsymbol{B}^{*} + \boldsymbol{K}_{1})\right.\right.\nonumber \\
& \qquad \left.\left.  -(\boldsymbol{B}^{*} + \boldsymbol{K}_{1}) \boldsymbol{K}^{-1}_{\Delta}(\boldsymbol{B}^{*} + \boldsymbol{K}_{2})     \right)         \right\}. \label{eq:M_1}
\end{align}

We are going to evaluate bounds of \eqref{eq:M_2} and \eqref{eq:M_1}, separately.

\subsubsection{Bounds of \eqref{eq:M_2}}
By applying Cram\'{e}r-Rao inequality of Lemma \ref{cri}, it can be shown that
\begin{align}
&J(X+W)^{-1} \nonumber \\
&\preceq \cov(X+W)=\cov(X)+ \boldsymbol{K}_{W} \\
& \overset{(a)}\preceq (\boldsymbol{B}^{*} + \boldsymbol{K}_{2}) \boldsymbol{K}^{-1}_{\Delta}(\boldsymbol{B}^{*} + \boldsymbol{K}_{1}) +\left(\boldsymbol{S}- \boldsymbol{B}^{*}\right),
\end{align}
where (a) is due to condition of $\cov(X) \preceq \boldsymbol{S}$ and \eqref{eq:KW3}.

Thus, \eqref{eq:M_2} can be bounded as
\begin{align}
&\tr \left\{  \boldsymbol{M}_{2} \left(      (\boldsymbol{B}^{*} + \boldsymbol{K}_{1}) \boldsymbol{K}^{-1}_{\Delta}(\boldsymbol{B}^{*} + \boldsymbol{K}_{2}) J(X+W)(\boldsymbol{B}^{*} + \boldsymbol{K}_{2}) \boldsymbol{K}^{-1}_{\Delta}(\boldsymbol{B}^{*} + \boldsymbol{K}_{1}) \right.\right.\nonumber \\
& \qquad \left.\left.  -(\boldsymbol{B}^{*} + \boldsymbol{K}_{1}) \boldsymbol{K}^{-1}_{\Delta}(\boldsymbol{B}^{*} + \boldsymbol{K}_{2})     \right)         \right\} \\
& \geq \tr \left\{    \boldsymbol{M}_{2}   (\boldsymbol{B}^{*} + \boldsymbol{K}_{1}) \boldsymbol{K}^{-1}_{\Delta}(\boldsymbol{B}^{*} + \boldsymbol{K}_{2})  \left(   (\boldsymbol{B}^{*} + \boldsymbol{K}_{2}) \boldsymbol{K}^{-1}_{\Delta}(\boldsymbol{B}^{*} + \boldsymbol{K}_{1}) +(\boldsymbol{S}- \boldsymbol{B}^{*})\right)^{-1} \right.\nonumber \\
& \quad \quad \left.\left( (\boldsymbol{B}^{*} + \boldsymbol{K}_{2}) \boldsymbol{K}^{-1}_{\Delta}(\boldsymbol{B}^{*} + \boldsymbol{K}_{1})      \right) -(\boldsymbol{B}^{*} + \boldsymbol{K}_{2}) \boldsymbol{K}^{-1}_{\Delta}(\boldsymbol{B}^{*} + \boldsymbol{K}_{1}) -\left(\boldsymbol{S}- \boldsymbol{B}^{*}\right)         \right\} \\
&= -\tr \left\{    \boldsymbol{M}_{2} \left(   (\boldsymbol{B}^{*} + \boldsymbol{K}_{1}) \boldsymbol{K}^{-1}_{\Delta}(\boldsymbol{B}^{*} + \boldsymbol{K}_{2})    (\boldsymbol{B}^{*} + \boldsymbol{K}_{2}) \boldsymbol{K}^{-1}_{\Delta}(\boldsymbol{B}^{*} + \boldsymbol{K}_{1}) +(\boldsymbol{S}- \boldsymbol{B}^{*})\right)^{-1}  \left(\boldsymbol{S}- \boldsymbol{B}^{*}\right)         \right\}\\
&\overset{(a)}= 0,\label{eq:01}
\end{align}
where (a) is from condition of $\left(\boldsymbol{S} - \boldsymbol{B}^{*} \right) \boldsymbol{M}_2 =0$.

\subsubsection{Bounds of \eqref{eq:M_1}}
By applying data processing inequality of Lemma \ref{DP_FI}, it can be seen that
\begin{align}
&J(X+W) \nonumber \\
&\preceq J(X+W|X)= \boldsymbol{K}^{-1}_{W} \\
&\overset{(a)}= \left((\boldsymbol{B}^{*} + \boldsymbol{K}_{2}) \boldsymbol{K}^{-1}_{\Delta}(\boldsymbol{B}^{*} + \boldsymbol{K}_{1}) - \boldsymbol{B}^{*}\right)^{-1},
\end{align}
where (a) is due to \eqref{eq:KW3}.

Thus, \eqref{eq:M_1} can be thereby bounded as
\begin{align}
&\tr \left\{  \boldsymbol{M}_{1} \left(      (\boldsymbol{B}^{*} + \boldsymbol{K}_{1}) \boldsymbol{K}^{-1}_{\Delta}(\boldsymbol{B}^{*} + \boldsymbol{K}_{2}) J(X+W)(\boldsymbol{B}^{*} + \boldsymbol{K}_{2}) \boldsymbol{K}^{-1}_{\Delta}(\boldsymbol{B}^{*} + \boldsymbol{K}_{1}) \right.\right.\nonumber \\
& \qquad \left.\left.  -(\boldsymbol{B}^{*} + \boldsymbol{K}_{1}) \boldsymbol{K}^{-1}_{\Delta}(\boldsymbol{B}^{*} + \boldsymbol{K}_{2})     \right)         \right\} \\
& \leq \tr \left\{    \boldsymbol{M}_{1}    (\boldsymbol{B}^{*} + \boldsymbol{K}_{1}) \boldsymbol{K}^{-1}_{\Delta}(\boldsymbol{B}^{*} + \boldsymbol{K}_{2})   \left( (\boldsymbol{B}^{*} + \boldsymbol{K}_{2}) \boldsymbol{K}^{-1}_{\Delta}(\boldsymbol{B}^{*} + \boldsymbol{K}_{1}) - \boldsymbol{B}^{*}\right)^{-1} \right.\nonumber \\
& \quad \quad \left.\left( (\boldsymbol{B}^{*} + \boldsymbol{K}_{2}) \boldsymbol{K}^{-1}_{\Delta}(\boldsymbol{B}^{*} + \boldsymbol{K}_{1})      -(\boldsymbol{B}^{*} + \boldsymbol{K}_{2}) \boldsymbol{K}^{-1}_{\Delta}(\boldsymbol{B}^{*} + \boldsymbol{K}_{1}) +\boldsymbol{B}^{*}  \right)       \right\} \\
& = \tr \left\{    \boldsymbol{M}_{1}    (\boldsymbol{B}^{*} + \boldsymbol{K}_{1}) \boldsymbol{K}^{-1}_{\Delta}(\boldsymbol{B}^{*} + \boldsymbol{K}_{2})   \left( (\boldsymbol{B}^{*} + \boldsymbol{K}_{2}) \boldsymbol{K}^{-1}_{\Delta}(\boldsymbol{B}^{*} + \boldsymbol{K}_{1}) - \boldsymbol{B}^{*}\right)^{-1} \boldsymbol{B}^{*}        \right\}\\
&\overset{(a)}=0, \label{eq:02}
\end{align}
where (a) is from condition of $\boldsymbol{B}^{*}  \boldsymbol{M}_1 =0$.

Combining \eqref{eq:01} and \eqref{eq:02}, we have shown $dg(\gamma)/d\gamma \geq 0$, for $\gamma \in (0,1)$. This completes the monotone path centered proof of Liu-Viswanath extremal inequality.

\section{Vector Generalization of Costa's Entropy Power Inequality}\label{sec3}
In \cite{LLPS13}, Liu \emph{et al.} derived another extremal inequality based on a generalized Costa's entropy power inequality, and use it to characterize secrecy capacity region of the MIMO Gaussian broadcast channel with layered confidential messages. In this section, we will proof the main theorem of \cite{LLPS13} by appealing to monotone path approach only.

\begin{theorem}\cite[Theroem 2]{LLPS13}\label{thm2}
Let $Z_{i}, i=0,\ldots,L$, be a total of $L+1$ Gaussian random vectors with covariance matrices $\boldsymbol{K}_{i}$, respectively. Assume that
\begin{equation}
\boldsymbol{K}_{1} \preceq \ldots \preceq \boldsymbol{K}_{L}.
\end{equation}
If there exists a positive semi-definite matrix such that
\begin{equation}
\sum_{i=1}^{L} \mu_{i} \left(   \boldsymbol{B}^{*} + \boldsymbol{K}_{i}        \right)^{-1} + \boldsymbol{M}_{1} = \left(   \boldsymbol{B}^{*} + \boldsymbol{K}_{0}        \right)^{-1} + \boldsymbol{M}_{2},\label{eq:57}
\end{equation}
for $\mu_{i} \geq 0$ with $\sum_{i=1}^{L} \mu _{i} = 1$, and $\boldsymbol{M}_1$, $\boldsymbol{M}_2$, $\boldsymbol{S}$ are positive semi-definite matrices satisfying
\begin{align}
\boldsymbol{B}^{*}\boldsymbol{M}_{1} = 0,\\
\left(\boldsymbol{S} - \boldsymbol{B}^{*} \right) \boldsymbol{M}_2 =0,
\end{align}
we have
\begin{equation}
\sum_{i=1}^{L}\mu_ih(X+Z_i) -h(X+Z_0)\leq \sum_{i=1}^{L}\frac{\mu_i}{2} \log \left|  (2 \pi e) \left(\boldsymbol{B}^{*} + \boldsymbol{K}_{i}\right)          \right| - \frac{1}{2} \log \left|  (2 \pi e)\left( \boldsymbol{B}^{*} + \boldsymbol{K}_{0}   \right)       \right|,\label{eq:main2}
\end{equation}
for any $X$ such that $\cov (X)\preceq \boldsymbol{S}$.
\end{theorem}
\begin{remark}
In \cite{LLPS13}, the extremal inequality is proved via enhancement argument and monotone path centered approach. However, as pointed out in \cite{CHW18}, the monotone path proof of \cite{LLPS13} contains an incorrect application of AM-GM inequality. We modified the perturbation framework, which is similar to Liu-Viswanath extremal inequality in the last section, and show the vector generalization of Costa's entropy power inequality can be proved without enhancement argument and  AM-GM inequality.
\end{remark}

\subsection{Monotone Path Construction}
We consider the same covariance preserved transform $\{X_{+,\gamma}, X_{-,\gamma}\}$ as \eqref{eq:trans1} and \eqref{eq:trans2}. For any $\gamma \in (0,1)$, define the perturbed function $g(\gamma)$ as follows
\begin{align}
g(\gamma) =& \sum_{i=1}^{L} \mu_{i} h\left(X_{+,\gamma}+\sqrt{1-\gamma}N_{i}+\sqrt{\gamma}N_{i}^{G}, X_{-,\gamma}+\sqrt{\gamma}N_{0}-\sqrt{1-\gamma}N^{G}_{0}\right),
\end{align}
where $(N_{0}^{G}, \ldots, N_{L}^{G})$ is a group of Gaussian random vectors with the same distribution as $(N_{0}, \ldots, N_{L})$, which is independent of $(N_{0}, \ldots, N_{L})$.

If $\gamma=0$, we have
\begin{align}
g(0) &=\sum_{i=1}^{L} \mu_{i} h(X+N_{i}, -X^{G}-N^{G}_{0})\\
     &=\sum_{i=1}^{L} \mu_{i} h(X+N_{i})+\frac{1}{2} \log \left|   (2 \pi e) \left(\boldsymbol{B}^{*}+\boldsymbol{K}_{0}\right)      \right|.
\end{align}
If $\gamma=1$, we have
\begin{align}
g(1) &=\sum_{i=1}^{L} \mu_{i} h(X^{G}+N^{G}_{i}, X+N_{0})\\
&=\sum_{i=1}^{L} \frac{\mu_{i}}{2} \log \left|   (2 \pi e) \left(\boldsymbol{B}^{*}+\boldsymbol{K}_{i}\right)      \right|+h(X+N_{0}).
\end{align}
Similarly, we are going to proof the monotonicity of $g(\gamma)$, \emph{i.e.},
\begin{equation}
\frac{d}{d\gamma} g(\gamma) \geq 0,   \quad \gamma \in (0,1).
\end{equation}

\subsection{Derivative Evaluation}

For the sake of simplifying notations, we denote by
\begin{align}
S_{i} &\triangleq \sqrt{\gamma}X^{G}+\sqrt{1-\gamma}N_{i} + \sqrt{\gamma} N^{G}_{i}, \quad i=1,\ldots, L, \label{eq:CS1}\\
S_{0} &\triangleq -\sqrt{1-\gamma}X^{G}+ \sqrt{\gamma}N_{0}- \sqrt{1-\gamma}N^{G}_{0}. \label{eq:CS2}
\end{align}

For $i=1,\ldots, L$, is can be evaluated that
\begin{align}
\sqrt{\gamma}S_{i} - \sqrt{1-\gamma}S_{0} &= X^{G} + \sqrt{\gamma(1-\gamma)} {N}_{i} + {\gamma}N^{G}_{i} - \sqrt{\gamma(1-\gamma)} {N}_{0}+ (1-\gamma)N^{G}_{0}.
\end{align}
and its covariance is shown as
\begin{align}
\boldsymbol{K}_{\Delta_{i}} &=\boldsymbol{B}^{*} + \gamma \boldsymbol{K}_{i} + (1-\gamma) \boldsymbol{K}_{0}. \label{eqn:CKW}
\end{align}

We can also represent $\sqrt{\gamma}S_i$ by using its optimal estimation when giving $\sqrt{\gamma}S_{i} - \sqrt{1-\gamma}S_{0}$, for $i=1,\ldots,L$,
\begin{align}
\sqrt{\gamma} S_i = \gamma (\boldsymbol{B}^{*} + \boldsymbol{K}_{i}) \boldsymbol{K}_{\Delta_{i}}^{-1}\left(    \sqrt{\gamma}S_{i} - \sqrt{1-\gamma}S_{0}  \right) + \sqrt{\gamma (1-\gamma)} W_{i} ,\label{eq:CW}
\end{align}
where $W_{i}$ is independent of $\sqrt{\gamma}S_{i} - \sqrt{1-\gamma}S_{0}$, with covariance
\begin{align}
\boldsymbol{K}_{W_{i}} &=\left(   {(1-\gamma)}(\boldsymbol{B}^{*} + \boldsymbol{K}_{i})^{-1} +  \gamma (\boldsymbol{B}^{*} + \boldsymbol{K}_{0})^{-1}           \right)^{-1} - \boldsymbol{B}^{*}\\
  &=(\boldsymbol{B}^{*} + \boldsymbol{K}_{0}) \boldsymbol{K}^{-1}_{\Delta_{i}}(\boldsymbol{B}^{*} + \boldsymbol{K}_{i}) - \boldsymbol{B}^{*}. \label{eq:CKW3}
\end{align}

Calculating the derivative of the bivariate differential entropy on $\gamma$ as \eqref{eq:ddev1}-\eqref{eq:dev1} similarly, we obtain
\begin{align}
& \frac{d}{d\gamma}g(\gamma)\nonumber \\
&=\sum_{i=1}^{L} \mu_{i} \frac{d}{d\gamma}h\left(X_{+,\gamma}+\sqrt{1-\gamma}N_{i}+\sqrt{\gamma}N_{i}^{G}, X_{-,\gamma}+\sqrt{\gamma}N_{0}-\sqrt{1-\gamma}N^{G}_{0}\right)\\
& = \sum_{i=1}^{L}\tr \left\{  \mu_{i} \left(      (\boldsymbol{B}^{*} + \boldsymbol{K}_{i})^{-1} - (\boldsymbol{B}^{*} + \boldsymbol{K}_{0})^{-1}  \right) \left(       (\boldsymbol{B}^{*} + \boldsymbol{K}_{i}) \boldsymbol{K}^{-1}_{\Delta_{i}}(\boldsymbol{B}^{*} + \boldsymbol{K}_{0}) J(X+W_i) \right.\right.\nonumber \\
& \qquad \qquad \qquad \qquad \left.\left.(\boldsymbol{B}^{*} + \boldsymbol{K}_{0}) \boldsymbol{K}^{-1}_{\Delta_{i}}(\boldsymbol{B}^{*} + \boldsymbol{K}_{i})  -(\boldsymbol{B}^{*} + \boldsymbol{K}_{i}) \boldsymbol{K}^{-1}_{\Delta_{i}}(\boldsymbol{B}^{*} + \boldsymbol{K}_{0})     \right)         \right\} \label{eq:low1} \\
&= \sum_{i=1}^{L}\tr \left\{  \mu_{i} \left(      (\boldsymbol{B}^{*} + \boldsymbol{K}_{i})^{-1} - (\boldsymbol{B}^{*} + \boldsymbol{K}_{0})^{-1} +\boldsymbol{M}_{1}-\boldsymbol{M}_{2}  \right) \left(       (\boldsymbol{B}^{*} + \boldsymbol{K}_{i}) \boldsymbol{K}^{-1}_{\Delta_{i}}(\boldsymbol{B}^{*} + \boldsymbol{K}_{0}) J(X+W_i) \right.\right.\nonumber \\
& \qquad \qquad \qquad \qquad \left.\left.(\boldsymbol{B}^{*} + \boldsymbol{K}_{0}) \boldsymbol{K}^{-1}_{\Delta_{i}}(\boldsymbol{B}^{*} + \boldsymbol{K}_{i})  -(\boldsymbol{B}^{*} + \boldsymbol{K}_{i}) \boldsymbol{K}^{-1}_{\Delta_{i}}(\boldsymbol{B}^{*} + \boldsymbol{K}_{0})     \right)         \right\} \label{eq:low2}\\
&\quad +\sum_{i=1}^{L} \tr \left\{ \mu_{i} \boldsymbol{M}_{2} \left(      (\boldsymbol{B}^{*} + \boldsymbol{K}_{i}) \boldsymbol{K}^{-1}_{\Delta_{i}}(\boldsymbol{B}^{*} + \boldsymbol{K}_{0}) J(X+W_i)(\boldsymbol{B}^{*} + \boldsymbol{K}_{0}) \boldsymbol{K}^{-1}_{\Delta_{i}}(\boldsymbol{B}^{*} + \boldsymbol{K}_{i}) \right.\right.\nonumber \\
&\quad \qquad \left.\left.  -(\boldsymbol{B}^{*} + \boldsymbol{K}_{i}) \boldsymbol{K}^{-1}_{\Delta_{i}}(\boldsymbol{B}^{*} + \boldsymbol{K}_{0})     \right)         \right\} \label{eq:low3}\\
& \quad - \sum_{i=1}^{L} \tr \left\{ \mu_{i} \boldsymbol{M}_{1} \left(      (\boldsymbol{B}^{*} + \boldsymbol{K}_{i}) \boldsymbol{K}^{-1}_{\Delta_{i}}(\boldsymbol{B}^{*} + \boldsymbol{K}_{0}) J(X+W_i)(\boldsymbol{B}^{*} + \boldsymbol{K}_{0}) \boldsymbol{K}^{-1}_{\Delta_{i}}(\boldsymbol{B}^{*} + \boldsymbol{K}_{i}) \right.\right.\nonumber \\
&\quad \qquad \left.\left.  -(\boldsymbol{B}^{*} + \boldsymbol{K}_{i}) \boldsymbol{K}^{-1}_{\Delta_{i}}(\boldsymbol{B}^{*} + \boldsymbol{K}_{0})     \right)         \right\}.\label{eq:low4}
\end{align}

\subsection{Lower Bounds}
It can be observed that \eqref{eq:low3} and \eqref{eq:low4} can be lower bounded by $0$, using the same method of \eqref{eq:M_1} and \eqref{eq:M_2}. We therefore omit the bounding details on \eqref{eq:low3} and \eqref{eq:low4}. In the following, we show that \eqref{eq:low2} is lower bounded by $0$.

Now let's denote
\begin{align}
\boldsymbol{I}_{i}&=(\boldsymbol{B}^{*} + \boldsymbol{K}_{i}) \boldsymbol{K}^{-1}_{\Delta_{i}}(\boldsymbol{B}^{*} + \boldsymbol{K}_{0}) J(X+W_i)(\boldsymbol{B}^{*} + \boldsymbol{K}_{0}) \boldsymbol{K}^{-1}_{\Delta_{i}}(\boldsymbol{B}^{*} + \boldsymbol{K}_{i})\nonumber \\
&\quad  -(\boldsymbol{B}^{*} + \boldsymbol{K}_{i}) \boldsymbol{K}^{-1}_{\Delta_{i}}(\boldsymbol{B}^{*} + \boldsymbol{K}_{0}), \quad i=1,\ldots,L.
\end{align}
For any $1 \leq j \leq i \leq L $, define
\begin{align}
&\boldsymbol{K}^{(j)}_{i} \triangleq \boldsymbol{K}_{W_i} - \boldsymbol{K}_{W_j} \\
&=(\boldsymbol{B}^{*} + \boldsymbol{K}_{0}) \boldsymbol{K}^{-1}_{\Delta_{i}}(\boldsymbol{B}^{*} + \boldsymbol{K}_{i}) - (\boldsymbol{B}^{*} + \boldsymbol{K}_{0}) \boldsymbol{K}^{-1}_{\Delta_{j}}(\boldsymbol{B}^{*} + \boldsymbol{K}_{j})\\
&=\left(   {(1-\gamma)}(\boldsymbol{B}^{*} + \boldsymbol{K}_{i})^{-1} +  \gamma (\boldsymbol{B}^{*} + \boldsymbol{K}_{0})^{-1}           \right)^{-1} - \left(   {(1-\gamma)}(\boldsymbol{B}^{*} + \boldsymbol{K}_{j})^{-1} +  \gamma (\boldsymbol{B}^{*} + \boldsymbol{K}_{0})^{-1}           \right)^{-1}\\
& \succeq 0.
\end{align}
Thus, $W_{i}$ can be decomposed as
\begin{equation}
W_{i} = W_{j} + W^{(j)}_{i},
\end{equation}
where $W^{(j)}_{i}$ is independent of $W_{j}$, which follows Gaussian distribution $\mathcal{N}(0, \boldsymbol{K}^{(j)}_{i})$.

Invoking Fisher information inequality of Lemma \ref{fi_inq} in Appendix \ref{app_lea2}, we can further obtain
\begin{align}
\boldsymbol{I}_{i} &= (\boldsymbol{B}^{*} + \boldsymbol{K}_{i}) \boldsymbol{K}^{-1}_{\Delta_{i}}(\boldsymbol{B}^{*} + \boldsymbol{K}_{0}) J(X+W_j+W^{(j)}_{i})(\boldsymbol{B}^{*} + \boldsymbol{K}_{0}) \boldsymbol{K}^{-1}_{\Delta_{i}}(\boldsymbol{B}^{*} + \boldsymbol{K}_{i})\nonumber \\
&\quad  -(\boldsymbol{B}^{*} + \boldsymbol{K}_{i}) \boldsymbol{K}^{-1}_{\Delta_{i}}(\boldsymbol{B}^{*} + \boldsymbol{K}_{0})\\
&\preceq (\boldsymbol{B}^{*} + \boldsymbol{K}_{j}) \boldsymbol{K}^{-1}_{\Delta_{j}}(\boldsymbol{B}^{*} + \boldsymbol{K}_{0}) J(X+W_j)(\boldsymbol{B}^{*} + \boldsymbol{K}_{0}) \boldsymbol{K}^{-1}_{\Delta_{j}}(\boldsymbol{B}^{*} + \boldsymbol{K}_{j})\nonumber \\
&\quad +\boldsymbol{K}^{(j)}_{i} -(\boldsymbol{B}^{*} + \boldsymbol{K}_{i}) \boldsymbol{K}^{-1}_{\Delta_{i}}(\boldsymbol{B}^{*} + \boldsymbol{K}_{0})\\
&=\boldsymbol{I}_{j}.
\end{align}
This implies the $\boldsymbol{I}_{i} \preceq\boldsymbol{I}_{j}$, for any $1 \leq j \leq i \leq L $.

For the sake of simplifying notations, we denote $\left\{ \boldsymbol{A}^{(1)}_{1}, \ldots,  \boldsymbol{A}^{(1)}_{L} \right\}$ as
\begin{align}
\boldsymbol{A}^{(1)}_{i} =  (\boldsymbol{B}^{*} + \boldsymbol{K}_{i})^{-1} - (\boldsymbol{B}^{*} + \boldsymbol{K}_{0})^{-1} +\boldsymbol{M}_{1}-\boldsymbol{M}_{2}, \quad i=1,\ldots,L.
\end{align}
 For $j=1,\ldots, L-1$, we recursively define $\left\{ \boldsymbol{A}^{(j+1)}_{j+1}, \ldots,  \boldsymbol{A}^{(j+1)}_{L} \right\}$ as
\begin{align}
\boldsymbol{A}^{(j+1)}_{i} = \boldsymbol{A}^{(j)}_{i} + \frac{\mu_{j}}{\sum_{k=j+1}^{L} \mu_{k}} \boldsymbol{A}^{(j)}_{j}, \quad i=j+1,\ldots,L. \label{eq:80}
\end{align}
We have the following proposition on sequence of $\left\{ \boldsymbol{A}^{(j)}_{j}, \ldots,  \boldsymbol{A}^{(j)}_{L} \right\}$, $j=1,\ldots,L$, and its proof can be found in Appendix \ref{proof_seq}.
\begin{proposition}\label{seq}
Let $\left\{ \boldsymbol{A}^{(j)}_{j}, \ldots,  \boldsymbol{A}^{(j)}_{L} \right\}$, $j=1,\dots,L$, we have the following properties,
\begin{enumerate}
\item [\emph{1)}]
\begin{equation}
\sum_{i=j}^{K} \mu_i \boldsymbol{A}^{(j)}_{i} = 0, \quad j=1, \ldots, L. \label{eq:81}
\end{equation}
Specifically, $\boldsymbol{A}^{(L)}_{L} = 0$.
\item [\emph{2)}]
\begin{equation}
\boldsymbol{A}^{(j)}_{j} \succeq 0. \label{eq:82}
\end{equation}
\end{enumerate}
\end{proposition}

We further define a family of functions $\left\{ f^{(1)}, \ldots, f^{(L)}\right\}$ as
\begin{align}
f^{(j)} = \sum_{i=j}^{L} \tr \left\{   \mu_{i}\boldsymbol{A}^{(j)}_{i}\boldsymbol{I}_{i}        \right\}, \quad j=1,\ldots,L.
\end{align}
Notice that $f^{(1)}$ is equivalent to \eqref{eq:low2}, and $f^{(L)}=0$. We are in the position of showing a chain of inequalities,
\begin{equation}
f^{(1)} \geq \cdots \geq f^{(L)}=0.
\end{equation}
 For a fixed $1\leq j \leq L-1$, we have
 \begin{align}
 f^{(j)} &= \sum_{i=j}^{L} \tr \left\{   \mu_{i}\boldsymbol{A}^{(j)}_{i}\boldsymbol{I}_{i}        \right\} \\
 &= \sum_{i=j+1}^{L} \tr \left\{ \mu_{i}\boldsymbol{A}^{(j)}_{i}\boldsymbol{I}_{i}       \right\}+ \tr \left\{ \mu_{j}\boldsymbol{A}^{(j)}_{j} \boldsymbol{I}_{j}\right\} \\
 &=\sum_{i=j+1}^{L} \tr \left\{ \mu_{i}  \left( \boldsymbol{A}^{(j)}_{i}\boldsymbol{I}_{i} +    \frac{\mu_{j}}{\sum_{i=j+1}^{L}\mu_i} \boldsymbol{A}^{(j)}_{j} \boldsymbol{I}_{j}  \right)  \right\} \\
 &\overset{(a)}\geq\sum_{i=j+1}^{L} \tr \left\{ \mu_{i}  \left( \boldsymbol{A}^{(j)}_{i}\boldsymbol{I}_{i} +    \frac{\mu_{j}}{\sum_{i=j+1}^{L}\mu_i} \boldsymbol{A}^{(j)}_{j} \boldsymbol{I}_{i}  \right)  \right\}\\
 & \overset{(b)} = \sum_{i=j+1}^{L}\tr \left\{ \mu_{i}  \boldsymbol{A}^{(j+1)}_{i}\boldsymbol{I}_{i}   \right\} \\
 &= f^{(j+1)},
 \end{align}
where
\begin{enumerate}
\item[(a)] is due to the fact that $\boldsymbol{I}_{i} \preceq \boldsymbol{I}_{j}$, for $1 \leq i \leq j \leq L$, and the fact that $\boldsymbol{A}_{j}^{(j)} \succeq 0$ in Proposition \ref{seq};
\item[(b)] is due to the recursive definition of $\boldsymbol{A}^{(j+1)}_{i}$ in \eqref{eq:80}.
\end{enumerate}
This completes the proof the inequality chain $f^{(1)} \geq \cdots \geq f^{(L)}=0$, and further shows that \eqref{eq:low2} is lower bounded by $0$.

\section{Capacity of the MIMO Gaussian Broadcast Channel with Common and Private Messages}\label{sec4}
\subsection{Problem Statement}
Consider the MIMO Gaussian broadcast channel defined as follows,
\begin{align}
Y_{1} = X + N_{1}, \\
Y_{2} =X + N_{2},
\end{align}
where ${X}$ is $p \times 1 $ dimensional input random vector, ${N}_{1}, {N}_{2}$ are $ p \times 1 $ dimensional additive noises, which follow Gaussian distribution $\mathcal{N}(0, \boldsymbol{K}_{1})$ and $\mathcal{N}(0, \boldsymbol{K}_{2})$. We may denote $n$-length \emph{i.i.d.} copies of random vector by a superscript $n$ in this section, \emph{e.g.} ${X}^{n} = \left( {X}_{1}, \ldots, {X}_{n}\right)$.\par

Assume one encoder wants to broadcast one common message $M_{0}$ to each receiver and two private messages  $M_{1}, M_{2}$ to every receiver respectively. We assume the message $ M_{0}, M_{1}, M_{2} $ are independent of each other and uniformly distributed over message sets $\mathcal{M}_{0}^{n} = \{ 1,2, \ldots, 2^{nR_{0}}\}$, $\mathcal{M}_{1}^{n} = \{ 1,2, \ldots, 2^{nR_{1}}\}$, $\mathcal{M}_{2}^{n} = \{ 1,2, \ldots, 2^{nR_{2}}\},$ individually.
A $(2^{nR_{0},2^{nR_{1}}, 2^{nR_{2}}},n)$ code consists of
\begin{itemize}
  \item an encoding function $\mathcal{\phi}^{n}: \mathcal{M}_{0}^{n} \times \mathcal{M}_{1}^{n} \times \mathcal{M}_{2}^{n} \mapsto \mathcal{R}^{p \times n}$ that assigns a codeword $x^{n}(m_{0},m_{1}, m_{2})$ to each message triple $(m_0, m_1,m_2)$,
  \item two decoding functions $\mathcal{\psi}_{i}^{n}: \mathcal{R}^{p \times n} \mapsto \mathcal{M}_{0}^{n} \times \mathcal{M}_{i}^{n}$, $i=1,2$, in which decoder $i$ assigns an estimate the common message and private message $(\hat{m}_{0i}, \hat{m}_{i})$ upon receiving $y^{n}_i$.
\end{itemize}
The average block error probability is defined as
\begin{equation}
P_{e}^{(n)} =  \max_{i= 1,2 }Pr \left\{  (\hat{M}_{0i}, \hat{M}_{i}) \neq  (M_{0}, M_{i}) \right\}.
\end{equation}
A rate triple $(R_{0}, R_{1}, R_{2})$ is said to be achievable if there exists a sequence of $( \mathcal{\phi}^{n}, \mathcal{\psi}^{n}_{1}, \mathcal{\psi}^{n}_{2}, n )$ such that $\lim_{n \rightarrow \infty} P_{e}^{(n)}= 0$.
\par

We assume an input covariance constraint  $ \cov(X) \preceq \boldsymbol{S} $ for some covariance matrix $\boldsymbol{S} \succeq 0 $. The capacity $\mathcal{C}(\boldsymbol{S})$ is the closure of all achievable rate triples. Since the capacity region is convex set, we can characterize the capacity by its supporting hyperplanes. Equivalently, it can be expressed by solving the following optimization problem\footnote{We omit the case that $R_{0}=0$, because it degenerates to the MIMO Gaussian broadcast channel with only private messages and has been resolved in \cite{WSS06,LV07}.}
\begin{equation} \label{core}
\max_{(R_{0},R_{1}, R_{2}) \in \mathcal{C}(\boldsymbol{S})} R_{0} + \mu_{1}R_{1} + \mu_{2}R_{2}.
\end{equation}

If $\max \{\mu_{1}, \mu_{2}\} \geq 1$, we have
\begin{align}
 \max_{(R_{0},R_{1}, R_{2}) \in\mathcal{C}(\boldsymbol{S})} R_{0} + \mu_{1}R_{1} + \mu_{2}R_{2}
\leq  \max_{(0 ,R_{1},R_{2}) \in \mathcal{C}(\boldsymbol{S})}  \mu_{1}R_{1} + \mu_{2}R_{2}. \label{eq:tmp}
\end{align}
Notice that the r.h.s. of \eqref{eq:tmp} indicates the capacity region of MIMO Gaussian broadcast channel with private messages only, which has been derived in\cite{WSS06,LV07}. It means an upper bound of the optimal value of \eqref{core} can be achieved by setting the rate of common message zero. So the capacity in the case that $\max \{\mu_{1}, \mu_{2} \} \geq 1$ is obtained by the capacity without common message. Throughout this section, we only focus on the case $0 < \mu_{1}, \mu_{2} < 1$.

\subsection{The Extremal Inequality}
By performing the coding scheme in \cite{JG04,WSS06-2,EU12,GN14}, an achievable rate region can be characterized by the matrix optimization problem
\begin{equation}
\max_{(R_{0},R_{1}, R_{2}) \in \mathcal{C}(\mathbf{S})} R_{0} + \mu_{2}R_{2} + \mu_{1} R_{1} \geq R^{*}(\boldsymbol{S}, \mu_1, \mu_2),  \quad 0 < \mu_{1} \leq \mu_{2}  < 1,
\end{equation}
where
\begin{align}
 & R^{*}(\boldsymbol{S}, \mu_1, \mu_2)\nonumber\\
 & \triangleq  \max_{\boldsymbol{B}_{1}, \boldsymbol{B}_{2}} \quad  \min_{i=1,2} \left\{ \frac{1}{2} \log \frac{|\boldsymbol{S}+ \boldsymbol{K}_{i}|}{|\boldsymbol{B}_{1} + \boldsymbol{B}_{2}+ \boldsymbol{K}_{i}|}\right\}
 +  \frac{\mu_{2}}{2} \log \frac{|\boldsymbol{B}_{1}  + \boldsymbol{B}_{2} + \boldsymbol{K}_{2}|}{|\boldsymbol{B}_{2} + \boldsymbol{K}_{2}|}
 + \frac{\mu_{1}}{2} \log \frac{|\boldsymbol{B}_{2}+\boldsymbol{K}_{1}|}{|\boldsymbol{K}_{1}|} ,\nonumber \\
& \quad\text{subject to} \quad  \boldsymbol{B}_{1} \succeq 0, \boldsymbol{B}_{2} \succeq 0, \nonumber\\
& \qquad\qquad \qquad   \boldsymbol{B}_{1}+\boldsymbol{B}_{2} \preceq \boldsymbol{S}.\label{optimal}
\end{align}

Let $\left(\boldsymbol{B}^{*}_{1}, \boldsymbol{B}^{*}_{2}\right)$ be a maximizer of $R^{*}(\boldsymbol{S}, \mu_1, \mu_2)$, the necessary Karush-Kuhn-Tucker (KKT) conditions are given in the following lemma, whose proof is omitted because it is a standard evaluation on Lagrange function of $R^{*}(\boldsymbol{S}, \mu_1, \mu_2)$.
\begin{lemma}\label{lemma_KKT1}
The maximizer $\left(\boldsymbol{B}^{*}_{1}, \boldsymbol{B}^{*}_{2}\right)$ of $R^{*}(\boldsymbol{S}, \mu_1, \mu_2)$ need to satisfy
\begin{align}
\frac{\mu_{2}-\lambda}{2}(\boldsymbol{B}_{1}^{*}  + \boldsymbol{B}_{2}^{*} + \boldsymbol{K}_{2})^{-1} + \boldsymbol{M}_{1}
=& \; \frac{1-\lambda}{2} (\boldsymbol{B}_{1}^{*} + \boldsymbol{B}_{2}^{*} + \boldsymbol{K}_{1})^{-1}+\boldsymbol{M}_{3},  \label{KKT_eq1}\\
\frac{\mu_{2}}{2} (\boldsymbol{B}_{2}^{*} + \boldsymbol{K}_{2})^{-1} + \boldsymbol{M}_{1}
=& \; \frac{\mu_{1}}{2} ( \boldsymbol{B}_{2}^{*} + \boldsymbol{K}_{1})^{-1} + \boldsymbol{M}_{2},\label{KKT_eq2}
\end{align}
for some positive semi-definite matrices $\boldsymbol{B}_{1}^{*}$, $\boldsymbol{B}_{2}^{*}$, $\boldsymbol{M}_{1}$, $\boldsymbol{M}_{2}$ and $\boldsymbol{M}_{3}$ such that
\begin{align}
\boldsymbol{B}_{1}^{*}\boldsymbol{M}_{1} &= 0,\\
\boldsymbol{B}_{2}^{*} \boldsymbol{M}_{2}&=0,\\
\left(\boldsymbol{S}-\boldsymbol{B}_{1}^{*}-\boldsymbol{B}_{2}^{*}\right)\boldsymbol{M}_{3}&=0.
\end{align}
\end{lemma}
To establish an extremal inequality and obtain the capacity region of two-receiver MIMO Gaussian broadcast channel common and private messages, we need further show
\begin{equation}
\max_{(R_{0},R_{1}, R_{2}) \in \mathcal{C}(\mathbf{S})} R_{0} + \mu_{2}R_{2} + \mu_{1} R_{1} \leq R^{*}(\boldsymbol{S}, \mu_1, \mu_2),
\end{equation}
for any $0 < \mu_{1} \leq \mu_{2}  < 1 $.

In \cite{GN14}, Geng and Nair resolved this converse proof of Gaussian optimality by exploiting the factorization of concave envelopes. Nevertheless, it seems indirect to extend this method to multi-receiver case. In this section, we construct the monotone path in tensorized probability space, and recover the same result as in \cite{GN14}. Thus, Our approach gives an alternative converse based on perturbation arguments such as \cite{LV07}.

Now consider the $UVW$ outer bound in \cite[Sec. III-B]{GN14}.
For any $(R_{0},R_{1}, R_{2}) \in \mathcal{C}(\boldsymbol{S}) $ and $0 <\mu_{1} \leq \mu_{2}\leq 1$, we have
\begin{align}
& R_{0} + \mu_{2}R_{2} + \mu_{1} R_{1} \nonumber \\
&\leq  (1-\mu_{2})R_{0} +  (\mu_{2} -\mu_{1})(R_{0}+R_{2})  + \mu_{1} (R_{0}+R_{2} +R_{1})  \\
&\leq  \min \{I(U;{Y}_{1}), I(U; {Y}_{2})\} + \mu_{2} I(V; {Y}_{2} | U) + \mu_{1}I({X}; {Y}_{1} | V, U)  \\
& \leq  \lambda I(U;{Y}_{2}) + (1-\lambda) I(U; {Y}_{1}) + \mu_{2} I(V; {Y}_{2} | U) + \mu_{1}I({X}; {Y}_{1} | V, U)  \\
&= \mu_{1}h({X}+{N}_{1} | U,V) - \mu_{2} h({X}+{N}_{2} | U,V) + (\mu_{2}- \lambda) h({X}+{N}_{2} | U) - (1-\lambda) h({X}+{N}_{1} | U) \nonumber \\
 &\quad - \mu_{1}h({N}_{1}) + \lambda h({X} + {N}_{2}) + (1-\lambda) h({X}+{N}_{1})  \\
&\leq  \mu_{1}h({X}+{N}_{1} | U,V) - \mu_{2} h({X}+{N}_{2} | U,V) + (\mu_{2}- \lambda) h({X}+{N}_{2} | U) - (1-\lambda) h({X}+{N}_{1} | U) \nonumber \\
 &\quad - \frac{\mu_{1}}{2} \log \left| (2\pi e) \boldsymbol{K}_{1} \right| + \frac{\lambda}{2} \log \left| (2\pi e) (\boldsymbol{S} + \boldsymbol{K}_{2}) \right| + \frac{1-\lambda}{2} \log \left| (2\pi e) (\boldsymbol{S} + \boldsymbol{K}_{1}) \right|. \label{eq:lb}
\end{align}
By comparing \eqref{eq:lb} with optimization problem $R^{*}(\boldsymbol{S}, {\mu_1, \mu_2})$ in \eqref{optimal}, it can be shown that to prove the converse part, it is sufficient to prove the following extremal inequality.

\begin{theorem} \label{mainthm3}
Let $\mathbf{B}_{1}^{*}$ and $\mathbf{B}_{2}^{*}$ be an optimal solution of $R^{*}(\boldsymbol{S}, {\mu_1, \mu_2})$, then for any random variables $(U,V, {X})$ such that $(U,V) \rightarrow {X} \rightarrow ({Y}_{1}, {Y}_{2})$ and $\cov (X) \preceq \boldsymbol{S}$, we have
\begin{align}
& \mu_{1}h({X}+{N}_{1} | U,V) - \mu_{2} h({X}+{N}_{2} | U,V) + (\mu_{2}- \lambda) h({X}+{N}_{2} | U) - (1-\lambda) h({X}+{N}_{1} | U) \nonumber \\
&  \leq  \frac{\mu_{1}}{2} \log \left| (2\pi e) (\boldsymbol{B}_{2}^{*} + \boldsymbol{K}_{1}) \right| - \frac{\mu_{2}}{2} \log \left| (2\pi e) (\boldsymbol{B}_{2}^{*} + \boldsymbol{K}_{2}) \right| + \frac{\mu_{2} - \lambda}{2} \log \left| (2\pi e) (\boldsymbol{B}_{1}^{*}+\boldsymbol{B}_{2}^{*} + \boldsymbol{K}_{2}) \right| \nonumber \\
 &\quad - \frac{1 - \lambda}{2} \log \left| (2\pi e) (\boldsymbol{B}_{1}^{*}+\boldsymbol{B}_{2}^{*} + \boldsymbol{K}_{1}) \right|.\label{eq:m3}
\end{align}
\end{theorem}

\subsection{Proof of Theorem \ref{mainthm3}}
\subsubsection{Monotone Path Construction} As before, we consider the covariance preserved transform $\{X^{(1)}_{+,\gamma}, X^{(1)}_{-,\gamma}, X^{(2)}_{+,\gamma}, X^{(2)}_{-,\gamma}\}$ as below,
\begin{align}
X^{(2)}_{+,\gamma} &=\sqrt{1-\gamma}X + \sqrt{\gamma}X_2^{G}, \label{eq:tr1}\\
X^{(2)}_{-,\gamma} &=\sqrt{\gamma}X - \sqrt{1-\gamma}X_2^{G}, \label{eq:tr2}\\
X^{(1)}_{+,\gamma} &=X^{(2)}_{+,\gamma}+\sqrt{\gamma}X_1^{G}, \label{eq:tr3}\\
X^{(1)}_{-,\gamma} &=X^{(2)}_{-,\gamma}- \sqrt{1-\gamma}X_1^{G}, \label{eq:tr4}
\end{align}
where $X_1^{G}$ and $X_{2}^{G}$ follow independent Gaussian distribution $\mathcal{N}(0, \boldsymbol{B}_{1}^{*})$ and $\mathcal{N}(0, \boldsymbol{B}_{2}^{*})$, and they are both independent of $X$.
For any $\gamma \in (0,1)$, define the perturbed function $g(\gamma)$ as
\begin{align}
g(\gamma) =& \mu_{2} h \left(\left. X^{(2)}_{+,\gamma}+\sqrt{1-\gamma}N_{1} + \sqrt{\gamma} N_{1}^{G}, X^{(2)}_{-,\gamma}+\sqrt{\gamma}N_{2} - \sqrt{1-\gamma} N_{2}^{G} \right| U,V\right)\nonumber \\
&-(\mu_{2}-\mu_{1}) h \left(\left. X^{(2)}_{+,\gamma}+\sqrt{1-\gamma}N_{1} + \sqrt{\gamma} N_{1}^{G} \right|U,V\right) \nonumber \\
&-(\mu_2-\lambda) h \left(\left. X^{(1)}_{+,\gamma}+\sqrt{1-\gamma}N_{1} + \sqrt{\gamma} N_{1}^{G}, X^{(1)}_{-,\gamma}+\sqrt{\gamma}N_{2} - \sqrt{1-\gamma} N_{2}^{G} \right| U \right)\nonumber \\
&-(1-\mu_{2})h \left(\left. X^{(1)}_{+,\gamma}+\sqrt{1-\gamma}N_{1} + \sqrt{\gamma} N_{1}^{G} \right|U \right),
\end{align}
where $N_{1}^{G}$, $N_{2}^{G}$ are Gaussian random vectors with the same distribution of $N_1$, $N_2$, which are independent of $N_1$, $N_2$.

When $\gamma=0$, we notice that
\begin{align}
g(0)& = \mu_2 h\left(\left.X+N_{1}, -X^{G}_{2} - N^{G}_{2} \right| U,V \right)- (\mu_2-\mu_1) h\left( X+N_1|U,V \right) \nonumber \\
&\quad - (\mu_2-\lambda) h \left( \left.X+N_{1}, -X^{G}_{1}-X^{G}_{2} - N^{G}_{2}\right|U \right) - (1-\mu_2)h \left( X+N_{1} |U    \right)\\
&=\mu_{1}h(X+N_{1}|U,V)-(1-\lambda)h(X+N_1|U) \nonumber \\
& \quad + \frac{\mu_{2}}{2} \log \left| (2\pi e) (\boldsymbol{B}_{2}^{*} + \boldsymbol{K}_{2}) \right|-\frac{\mu_{2} - \lambda}{2} \log \left| (2\pi e) (\boldsymbol{B}_{1}^{*}+\boldsymbol{B}_{2}^{*} + \boldsymbol{K}_{2}) \right|.
\end{align}
When $\gamma=1$, we notice that
\begin{align}
g(1) &=\mu_2 h \left( \left. X^{G}_{2}+N^{G}_{2},X+N_{2} \right| U,V      \right) - (\mu_2-\mu_1) h\left(X^{G}_{2}+N^{G}_{2}\right)\nonumber \\
      &\quad -(\mu_2-\lambda) h \left(\left.   X^{G}_{1}+X^{G}_{2}+N^{G}_{1},       X+N_{2}        \right|U  \right) - (1-\mu_2)h \left(  X^{G}_{1}+X^{G}_{2}+N^{G}_{1}       \right) \\
      &= \mu_{2} h(X+N_{2}|U,V) - (\mu_2-\lambda) h(X+N_{2}|U) \nonumber \\
      &\quad + \frac{\mu_{1}}{2} \log \left| (2\pi e) (\boldsymbol{B}_{2}^{*} + \boldsymbol{K}_{1}) \right|- \frac{1 - \lambda}{2} \log \left| (2\pi e) (\boldsymbol{B}_{1}^{*}+\boldsymbol{B}_{2}^{*} + \boldsymbol{K}_{1}) \right|.
\end{align}

To prove \eqref{eq:m3} in Theorem \ref{mainthm3}, it is sufficient to show that $g(\gamma)$ is monotonically increasing along the path of $\{X^{(1)}_{+,\gamma}, X^{(1)}_{-,\gamma}, X^{(2)}_{+,\gamma}, X^{(2)}_{-,\gamma}\}$, \emph{i.e.},
\begin{equation}
\frac{d}{d\gamma} g(\gamma) \geq 0,   \quad \gamma \in (0,1).
\end{equation}

\subsubsection{Derivative Evaluation}
We firstly denote by
\begin{align}
S^{(2)}_{1} &\triangleq \sqrt{\gamma}X_{2}^{G}+\sqrt{1-\gamma}N_{1} + \sqrt{\gamma} N^{G}_{1}, \label{eq:S1_2}\\
S^{(2)}_{2} &\triangleq -\sqrt{1-\gamma}X_{2}^{G}+ \sqrt{\gamma}N_{2}- \sqrt{1-\gamma}N^{G}_{2}. \label{eq:S2_2}
\end{align}
Their covariance are
\begin{align}
\boldsymbol{K}_{S^{(2)}_1} &= \gamma \boldsymbol{B}^{*}_{2} + \boldsymbol{K}_{1},\\
\boldsymbol{K}_{S^{(2)}_2} &= (1-\gamma) \boldsymbol{B}^{*}_{2} + \boldsymbol{K}_{2}.
\end{align}

Using the same method in \eqref{eq:ddev1}-\eqref{eq:dev1}, the derivative of first term in $g(\gamma)$ can be written as
\begin{align}
&\frac{d}{d\gamma} h \left(\left. X^{(2)}_{+,\gamma}+\sqrt{1-\gamma}N_{1} + \sqrt{\gamma} N_{1}^{G}, X^{(2)}_{-,\gamma}+\sqrt{\gamma}N_{2} - \sqrt{1-\gamma} N_{2}^{G} \right| U,V\right) \nonumber \\
&=\frac{d}{d\gamma} h\left( \left. \sqrt{1-\gamma}X+S^{(2)}_{1} ,  \sqrt{\gamma}X+S^{(2)}_{2} \right| U,V     \right) \\
& =\tr \left\{  \left(     (\boldsymbol{B}_{2}^{*} + \boldsymbol{K}_{1})^{-1} - (\boldsymbol{B}_{2}^{*} + \boldsymbol{K}_{2})^{-1}  \right) \left(   (\boldsymbol{B}_{2}^{*} + \boldsymbol{K}_{1}) \boldsymbol{K}^{-1}_{\Delta^{(2)}}(\boldsymbol{B}_{2}^{*} + \boldsymbol{K}_{2}) J(X+W| U,V) \right.\right.\nonumber \\
& \qquad \qquad \left.\left. (\boldsymbol{B}_{2}^{*} + \boldsymbol{K}_{2}) \boldsymbol{K}^{-1}_{\Delta^{(2)}}(\boldsymbol{B}_{2}^{*} + \boldsymbol{K}_{1}) -(\boldsymbol{B}_{2}^{*} + \boldsymbol{K}_{1}) \boldsymbol{K}^{-1}_{\Delta^{(2)}}(\boldsymbol{B}_{2}^{*} + \boldsymbol{K}_{2})     \right)         \right\},\label{eqn:subto_2}
\end{align}
where
\begin{equation}
W=\sqrt{1-\gamma} S^{(2)}_2 - \mathbb{E} \left[ \sqrt{1-\gamma} S^{(2)}_2 \left| \sqrt{\gamma} S^{(2)}_{1}-\sqrt{1-\gamma} S^{(2)}_2    \right.   \right],\label{eqn:W_2}
\end{equation}
\begin{equation}
\boldsymbol{K}_{\Delta^{(2)}} = \boldsymbol{B}_{2}^{*} + \gamma \boldsymbol{K}_{1} + (1-\gamma) \boldsymbol{K}_{2}. \label{eqn:KW2}
\end{equation}

We now represent $\sqrt{\gamma}S^{(2)}_1$ by using its optimal estimation when giving $\sqrt{1-\gamma}S^{(2)}_2$,
\begin{align}
\sqrt{\gamma}S^{(2)}_1 = -\gamma \boldsymbol{B}_{2}^{*} \boldsymbol{K}_{S^{(2)}_2}^{-1} \sqrt{1-\gamma}S^{(2)}_{2} + \left(  \boldsymbol{B}_{2}^{*} + \boldsymbol{K}_{2}         \right)\boldsymbol{K}_{S^{(2)}_2}^{-1} W',\label{eqn:S1_2}
\end{align}
where $W'$ is independent of $S^{(2)}_{2}$, which is a Gaussian random vector with covariance $\boldsymbol{K}_{W'}$.
Since
\begin{align}
\sqrt{\gamma}S^{(2)}_1 -  \sqrt{1-\gamma}S^{(2)}_{2} = \left(  \boldsymbol{B}_{2}^{*} + \boldsymbol{K}_{2}         \right)\boldsymbol{K}_{S^{(2)}_2}^{-1} \left( -\sqrt{1-\gamma}S^{(2)}_{2} + W'\right),\label{eqn:W'_2}
\end{align}
we have the following relation on covariance matrices,
\begin{align}
\boldsymbol{K}_{\Delta^{(2)}} = \left(  \boldsymbol{B}_2^{*} + \boldsymbol{K}_{2}         \right)\boldsymbol{K}_{S^{(2)}_2}^{-1} \left(     (1-\gamma)\boldsymbol{K}_{S^{(2)}_2} +\boldsymbol{\Sigma}_{W'}    \right)\boldsymbol{K}_{S^{(2)}_2}^{-1}\left(  \boldsymbol{B}^{*} + \boldsymbol{\Sigma}_{2}         \right). \label{eqn:WW_2}
\end{align}
From \eqref{eqn:W_2}, we have
\begin{align}
 \sqrt{\gamma (1-\gamma)} W &= \sqrt{1-\gamma} S^{(2)}_2 +(1-\gamma) (\boldsymbol{B}^{*} + \boldsymbol{K}_{2}) \boldsymbol{\Sigma}_{\Delta^{(2)}}^{-1}\left(    \sqrt{\gamma}S^{(2)}_{1} - \sqrt{1-\gamma}S^{(2)}_{2}  \right) \\
 &\overset{(a)}=\sqrt{1-\gamma} S^{(2)}_2 +(1-\gamma) (\boldsymbol{B}_2^{*} + \boldsymbol{K}_{2}) \boldsymbol{K}_{\Delta^{(2)}}^{-1}\left(  \boldsymbol{B}_2^{*} + \boldsymbol{K}_{2}         \right)\boldsymbol{K}^{-1}_{S^{(2)}_2} \left( -\sqrt{1-\gamma}S^{(2)}_{2} + W'\right) \\
 &\overset{(b)} = \sqrt{1-\gamma} S^{(2)}_2 +(1-\gamma)\boldsymbol{K}_{S^{(2)}_2} \left(     (1-\gamma) \boldsymbol{K}_{S^{(2)}_2} + \boldsymbol{\Sigma}_{W'}    \right)^{-1}\left( -\sqrt{1-\gamma}S^{(2)}_{2} + W'\right)\\
 &=\boldsymbol{\Sigma}_{W'} \left(     (1-\gamma) \boldsymbol{\Sigma}_{S^{(2)}_2} + \boldsymbol{\Sigma}_{W'}    \right)^{-1}\sqrt{1-\gamma}S^{(2)}_{2} + (1-\gamma)\boldsymbol{\Sigma}_{S^{(2)}_2} \left(     (1-\gamma)\boldsymbol{K}_{S_2^{(2)}} + \boldsymbol{\Sigma}_{W'}    \right)^{-1}W',
\end{align}
where (a) is due to \eqref{eqn:W'_2}, and (b) is due to \eqref{eqn:WW_2}.

Invoking the complementary identity in Corollary \ref{comp} of Appendix \ref{app_lea2}, we write Fisher information $J(X+W|U,V)$ as
\begin{align}
&J(X+W|U,V) \nonumber \\
&=\gamma(1-\gamma)J\left(\left.\sqrt{\gamma (1-\gamma)} X + \sqrt{\gamma (1-\gamma)} W \right|U,V\right)\\
&=\frac{\gamma}{1-\gamma}\boldsymbol{K}_{S^{(2)}_2}^{-1}((1-\gamma)\boldsymbol{K}_{S^{(2)}_2} + \boldsymbol{K}_{W'})J \left(\left. \sqrt{\gamma (1-\gamma)} X+W'\right| \sqrt{\gamma(1-\gamma)}X + \sqrt{1-\gamma}S^{(2)}_{2},U,V \right)\nonumber \\
&\quad ((1-\gamma)\boldsymbol{K}_{S^{(2)}_2} + \boldsymbol{K}_{W'})\boldsymbol{K}_{S^{(2)}_2}^{-1} -\frac{\gamma}{1-\gamma}\boldsymbol{K}_{S^{(2)}_2}^{-1}((1-\gamma)\boldsymbol{K}_{S^{(2)}_2} + \boldsymbol{K}_{W'})\boldsymbol{K}_{S^{(2)}_2}^{-1}\label{eqn:su_1}\\
&\overset{(a)}=\frac{\gamma}{1-\gamma}\boldsymbol{K}_{S^{(2)}_2}^{-1}((1-\gamma)\boldsymbol{K}_{S^{(2)}_2} + \boldsymbol{K}_{W'})\boldsymbol{K}^{-1}_{S^{(2)}_2}\left(  \boldsymbol{B}_{2}^{*} + \boldsymbol{K}_{2}         \right)J \left(\left. \sqrt{\gamma (1-\gamma)} X+\sqrt{\gamma}S^{(2)}_{1}\right| \sqrt{\gamma(1-\gamma)}X + \sqrt{1-\gamma}S^{(2)}_{2}, U,V \right)\nonumber \\
&\quad\left(  \boldsymbol{B}_{2}^{*} + \boldsymbol{K}_{2}         \right)\boldsymbol{K}^{-1}_{S^{(2)}_2} ((1-\gamma)\boldsymbol{K}_{S^{(2)}_2} + \boldsymbol{K}_{W'}) -\frac{\gamma}{1-\gamma}\boldsymbol{K}_{S^{(2)}_2}^{-1}((1-\gamma)\boldsymbol{K}_{S^{(2)}_2} + \boldsymbol{K}_{W'})\boldsymbol{K}_{S^{(2)}_2}^{-1} \\
&\overset{(b)} = \frac{1}{1-\gamma}(\boldsymbol{B}_{2}^{*} + \boldsymbol{K}_{2})^{-1}\boldsymbol{K}_{\Delta^{(2)}} J\left(\left. \sqrt{1-\gamma} X+S^{(2)}_{1}\right| \sqrt{\gamma}X + S^{(2)}_{2}, U,V \right)\boldsymbol{K}_{\Delta^{(2)}}(\boldsymbol{B}_{2}^{*} + \boldsymbol{K}^{(2)}_{2})^{-1}\nonumber \\
& \quad-\frac{\gamma}{1-\gamma}(\boldsymbol{B}_{2}^{*} + \boldsymbol{K}_{2})^{-1}\boldsymbol{K}_{\Delta^{(2)}}(\boldsymbol{B}_{2}^{*} + \boldsymbol{K}_{2})^{-1},\label{eqn:su_2}
\end{align}
where (a) is due to \eqref{eqn:S1_2}, and (b) is due to \eqref{eqn:WW_2}.

Substituting \eqref{eqn:su_2} into \eqref{eqn:subto_2}, we obtain
\begin{align}
& \frac{d}{d\gamma} h\left( \left. \sqrt{1-\gamma}X+S^{(2)}_{1} ,  \sqrt{\gamma}X+S^{(2)}_{2} \right| U,V     \right)\nonumber \\
& =\frac{1}{2(1-\gamma)}\tr \left\{    \left(  \boldsymbol{B}_2^{*} + \boldsymbol{K}_{1}     \right)  \left(     (\boldsymbol{B}_2^{*} + \boldsymbol{K}_{1})^{-1} - \left(  \boldsymbol{B}_2^{*} + \boldsymbol{K}_{2}     \right)^{-1}       \right)   \left(  \boldsymbol{B}_2^{*} + \boldsymbol{K}_{1}     \right) J   \left(  \sqrt{1-\gamma}X+S^{(2)}_{1} \left|   \sqrt{\gamma}X+S^{(2)}_{2}, U,V \right. \right)         \right\} \nonumber \\
&\quad -\frac{1}{2(1-\gamma)}\tr \left\{  \gamma \left(  \boldsymbol{B}_2^{*} + \boldsymbol{K}_{1}     \right)  \left(     (\boldsymbol{B}_2^{*} + \boldsymbol{K}_{1})^{-1} - \left(  \boldsymbol{B}_2^{*} + \boldsymbol{K}_{2}     \right)^{-1}       \right)   \left(  \boldsymbol{B}_2^{*} + \boldsymbol{K}_{1}     \right)      \boldsymbol{K}_{\Delta^{(2)}}^{-1}                   \right\} \nonumber \\
&\quad + \frac{1}{2(1-\gamma)}\tr \left\{  (1-\gamma) \left(  \boldsymbol{B}_2^{*} + \boldsymbol{K}_{2}     \right)  \left(     (\boldsymbol{B}_2^{*} -\boldsymbol{K}_{1})^{-1} - \left(  \boldsymbol{B}_2^{*} + \boldsymbol{K}_{2}     \right)^{-1}       \right)   \left(  \boldsymbol{B}_2^{*} + \boldsymbol{K}_{1}     \right)      \boldsymbol{K}_{\Delta^{(2)}}^{-1}                   \right\}\\
&=\frac{1}{2(1-\gamma)}\tr \left\{     \left(     (\boldsymbol{B}_2^{*} + \boldsymbol{K}_{1})^{-1} - \left(  \boldsymbol{B}_2^{*} + \boldsymbol{K}_{2}     \right)^{-1}       \right) \left(  \left(  \boldsymbol{B}_2^{*} + \boldsymbol{K}_{1}     \right)\right.\right. \nonumber \\
& \hspace{1in}\left.\left.     J   \left(  \sqrt{1-\gamma}X+S^{(2)}_{1} \left|   \sqrt{\gamma}X+S^{(2)}_{2}, U,V \right. \right) \left(  \boldsymbol{B}_2^{*} + \boldsymbol{K}_{1}     \right)  -\left(  \boldsymbol{B}_2^{*} + \boldsymbol{K}_{1}     \right)\right)     \right\}.\label{eq:tem1}
\end{align}

Using the same method as \eqref{eq:ddev2}-\eqref{eq:ddevv2}, the derivative of second term in $g(\gamma)$ can be evaluated as
\begin{align}
& \frac{d}{d\gamma} h\left( \left. \sqrt{1-\gamma}X+S^{(2)}_{1}  \right| U,V     \right)\nonumber \\
&= \frac{1}{2(1-\gamma)}\tr \left\{         (\boldsymbol{B}_2^{*} + \boldsymbol{K}_{1})^{-1}         \left(  \left(  \boldsymbol{B}_2^{*} + \boldsymbol{K}_{1}     \right) J   \left(\left.  \sqrt{1-\gamma}X+S^{(2)}_{1} \right|   U,V  \right)\left(  \boldsymbol{B}_2^{*} + \boldsymbol{K}_{1}     \right)  -\left(  \boldsymbol{B}_2^{*} + \boldsymbol{K}_{1}     \right)\right)     \right\}.\label{eq:tem2}
\end{align}
Similarly, the third and forth terms in $g(\gamma)$ can be evaluated as
\begin{align}
&\frac{d}{d\gamma} h\left( \left. \sqrt{1-\gamma}X+\sqrt{\gamma}X^{G}_{1}+S^{(2)}_{1} ,  \sqrt{\gamma}X-\sqrt{1-\gamma}X^{G}_{1}+S^{(2)}_{2} \right| U     \right)\nonumber \\
&=\frac{1}{2(1-\gamma)}\tr \left\{     \left(     (\boldsymbol{B}_2^{*}+\boldsymbol{B}_2^{*} + \boldsymbol{K}_{1})^{-1} - \left(  \boldsymbol{B}_1^{*}+\boldsymbol{B}_2^{*} + \boldsymbol{K}_{2}     \right)^{-1}       \right) \left(  \left( \boldsymbol{B}_1^{*}+ \boldsymbol{B}_2^{*} + \boldsymbol{K}_{1}     \right) \right.\right. \nonumber \\
& \hspace{1in}\left.\left.    J   \left(  \sqrt{1-\gamma}X+\sqrt{\gamma}X^{G}_{1}+S^{(2)}_{1} \left|   \sqrt{\gamma}X-\sqrt{1-\gamma}X_{1}^{G}+S^{(2)}_{2}, U \right. \right) \left( \boldsymbol{B}_1^{*} +\boldsymbol{B}_2^{*} + \boldsymbol{K}_{1}     \right)  -\left( \boldsymbol{B}_1^{*}+ \boldsymbol{B}_2^{*} + \boldsymbol{K}_{1}     \right)\right)     \right\},\label{eq:tem3}\\
&\frac{d}{d\gamma} h\left( \left. \sqrt{1-\gamma}X+\sqrt{\gamma}X^{G}_{1}+S^{(2)}_{1}  \right| U     \right)\nonumber \\
&=\frac{1}{2(1-\gamma)}\tr \left\{     \left(     (\boldsymbol{B}_1^{*}+\boldsymbol{B}_2^{*} + \boldsymbol{K}_{1})^{-1}     \right) \left(  \left(  \boldsymbol{B}_1^{*}+\boldsymbol{B}_2^{*} + \boldsymbol{K}_{1}     \right) \right.\right. \nonumber \\
& \hspace{1in} \left.\left. J   \left( \left. \sqrt{1-\gamma}X+\sqrt{\gamma}X^{G}_{1}+S^{(2)}_{1} \right|  U  \right) \left( \boldsymbol{B}_1^{*}+ \boldsymbol{B}_2^{*} + \boldsymbol{K}_{1}     \right)  -\left( \boldsymbol{B}_1^{*}+ \boldsymbol{B}_2^{*} + \boldsymbol{K}_{1}     \right)\right)     \right\}. \label{eq:tem4}
\end{align}

combining \eqref{eq:tem1}, \eqref{eq:tem2}, \eqref{eq:tem3} and \eqref{eq:tem4}, we obtain
\begin{align}
& 2(1-\gamma)\frac{d}{d\gamma}g(\gamma) \nonumber \\
&=\tr \left\{    \mu_2 \left(     (\boldsymbol{B}_2^{*} + \boldsymbol{K}_{1})^{-1} - \left(  \boldsymbol{B}_2^{*} + \boldsymbol{K}_{2}     \right)^{-1}       \right) \left(  \left(  \boldsymbol{B}_2^{*} + \boldsymbol{K}_{1}     \right) \right.\right. \nonumber \\
& \quad\qquad\left.\left.   J   \left(  \sqrt{1-\gamma}X+S^{(2)}_{1} \left|   \sqrt{\gamma}X+S^{(2)}_{2}, U,V \right. \right)  \left(  \boldsymbol{B}_2^{*} + \boldsymbol{K}_{1}     \right)  -\left(  \boldsymbol{B}_2^{*} + \boldsymbol{K}_{1}     \right)\right)     \right\}  \\
&  - \tr \left\{       (\mu_2-\mu_1)   (\boldsymbol{B}_2^{*} + \boldsymbol{K}_{1})^{-1}  \left(  \left(  \boldsymbol{B}_2^{*} + \boldsymbol{K}_{1}     \right) J   \left(\left.  \sqrt{1-\gamma}X+S^{(2)}_{1} \right|   U,V  \right)\left(  \boldsymbol{B}_2^{*} + \boldsymbol{K}_{1}     \right)  -\left(  \boldsymbol{B}_2^{*} + \boldsymbol{K}_{1}     \right)\right)     \right\} \\
&-\tr \left\{     \left(  (\mu_2-\lambda)   (\boldsymbol{B}_1^{*}+\boldsymbol{B}_2^{*} + \boldsymbol{K}_{1})^{-1} - \left( \boldsymbol{B}_1^{*}+ \boldsymbol{B}_2^{*} + \boldsymbol{K}_{2}     \right)^{-1}       \right) \left(  \left( \boldsymbol{B}_1^{*}+ \boldsymbol{B}_2^{*} + \boldsymbol{K}_{1}     \right) \right.\right. \nonumber \\
& \quad\qquad\left.\left.    J   \left(  \sqrt{1-\gamma}X+\sqrt{\gamma}X^{G}_{1}+S^{(2)}_{1} \left|   \sqrt{\gamma}X-\sqrt{1-\gamma}X_{1}^{G}+S^{(2)}_{2}, U \right. \right) \left(  \boldsymbol{B}_1^{*}+\boldsymbol{B}_2^{*} + \boldsymbol{K}_{1}     \right)  -\left( \boldsymbol{B}_1^{*}+ \boldsymbol{B}_2^{*} + \boldsymbol{K}_{1}     \right)\right)     \right\},\\
&-\tr \left\{     \left(   (1-\mu_2)  (\boldsymbol{B}_1^{*}+\boldsymbol{B}_2^{*} + \boldsymbol{K}_{1})^{-1}  \right) \left(  \left( \boldsymbol{B}_1^{*}+ \boldsymbol{B}_2^{*} + \boldsymbol{K}_{1}     \right) \right.\right. \nonumber \\
& \quad\qquad\left.\left.    J   \left( \left. \sqrt{1-\gamma}X+\sqrt{\gamma}X^{G}_{1}+S^{(2)}_{1} \right|  U  \right) \left( \boldsymbol{B}_1^{*}+ \boldsymbol{B}_2^{*} + \boldsymbol{K}_{1}     \right)  -\left(  \boldsymbol{B}_1^{*}+\boldsymbol{B}_2^{*} + \boldsymbol{K}_{1}     \right)\right)     \right\}.
\end{align}

\subsubsection{Lower Bounds}
Firstly applying data processing inequality in Lemma \ref{DP_FI},
\begin{align}
&J   \left(\left.  \sqrt{1-\gamma}X+S^{(2)}_{1} \right|   U,V  \right) \preceq J   \left(\left.  \sqrt{1-\gamma}X+S^{(2)}_{1} \right| \sqrt{\gamma}X+S^{(2)}_2,  U,V  \right),
\end{align}
\begin{align}
 &  J   \left( \left. \sqrt{1-\gamma}X+\sqrt{\gamma}X^{G}_{1}+S^{(2)}_{1} \right|  U  \right)
  \preceq J   \left( \left. \sqrt{1-\gamma}X+\sqrt{\gamma}X^{G}_{1}+S^{(2)}_{1} \right| \sqrt{\gamma}X-\sqrt{1-\gamma}X^{G}_{1}+S^{(2)}_{2},  U  \right).
\end{align}
We thus have
\begin{align}
& 2(1-\gamma)\frac{d}{d\gamma}g(\gamma) \nonumber \\
& \geq \tr \left\{     \left(  \mu_1   (\boldsymbol{B}_2^{*} + \boldsymbol{K}_{1})^{-1} -\mu_2 \left(  \boldsymbol{B}_2^{*} + \boldsymbol{K}_{2}     \right)^{-1}       \right) \left(  \left(  \boldsymbol{B}_2^{*} + \boldsymbol{K}_{1}     \right) \right.\right. \nonumber \\
& \quad\qquad\left.\left.   J   \left(  \sqrt{1-\gamma}X+S^{(2)}_{1} \left|   \sqrt{\gamma}X+S^{(2)}_{2}, U,V \right. \right)  \left(  \boldsymbol{B}_2^{*} + \boldsymbol{K}_{1}     \right)  -\left(  \boldsymbol{B}_2^{*} + \boldsymbol{K}_{1}     \right)\right)     \right\}  \\
&-\tr \left\{     \left(  (1-\lambda)   (\boldsymbol{B}_1^{*}+\boldsymbol{B}_2^{*} + \boldsymbol{K}_{1})^{-1} -(\mu_2-\lambda) \left( \boldsymbol{B}_1^{*}+ \boldsymbol{B}_2^{*} + \boldsymbol{K}_{2}     \right)^{-1}       \right) \left(  \left( \boldsymbol{B}_1^{*}+ \boldsymbol{B}_2^{*} + \boldsymbol{K}_{1}     \right) \right.\right. \nonumber \\
& \quad\qquad\left.\left.    J   \left(  \sqrt{1-\gamma}X+\sqrt{\gamma}X^{G}_{1}+S^{(2)}_{1} \left|   \sqrt{\gamma}X-\sqrt{1-\gamma}X_{1}^{G}+S^{(2)}_{2}, U \right. \right) \left(  \boldsymbol{B}_1^{*}+\boldsymbol{B}_2^{*} + \boldsymbol{K}_{1}     \right)  -\left( \boldsymbol{B}_1^{*}+ \boldsymbol{B}_2^{*} + \boldsymbol{K}_{1}     \right)\right)     \right\}\\
&\overset{(a)}= -\tr \left\{ \boldsymbol{M}_1   \left(  \left( \boldsymbol{B}_1^{*}+ \boldsymbol{B}_2^{*} + \boldsymbol{K}_{1}     \right) J   \left(  \sqrt{1-\gamma}X+\sqrt{\gamma}X^{G}_{1}+S^{(2)}_{1} \left|   \sqrt{\gamma}X-\sqrt{1-\gamma}X_{1}^{G}+S^{(2)}_{2}, U \right. \right)    \left( \boldsymbol{B}_1^{*}+ \boldsymbol{B}_2^{*} + \boldsymbol{K}_{1}     \right) \right.\right. \nonumber \\
&\qquad \qquad\left.\left. -   \left( \boldsymbol{B}_2^{*} + \boldsymbol{K}_{1}     \right) J   \left(  \sqrt{1-\gamma}X+S^{(2)}_{1} \left|   \sqrt{\gamma}X+S^{(2)}_{2}, U,V \right. \right)    \left(  \boldsymbol{B}_2^{*} + \boldsymbol{K}_{1}     \right) - \boldsymbol{B}_1^{*}                                   \right)\right\}                            \label{eq:term1} \\
&\quad-\tr \left\{     \boldsymbol{M}_2 \left(  \left(  \boldsymbol{B}_2^{*} + \boldsymbol{K}_{1}     \right) J   \left(  \sqrt{1-\gamma}X+S^{(2)}_{1} \left|   \sqrt{\gamma}X+S^{(2)}_{2}, U,V \right. \right)  \left(  \boldsymbol{B}_2^{*} + \boldsymbol{K}_{1}     \right)  -\left(  \boldsymbol{B}_2^{*} + \boldsymbol{K}_{1}     \right)\right)     \right\} \label{eq:term2} \\
&\quad +\tr \left\{     \boldsymbol{M}_{3} \left(  \left( \boldsymbol{B}_1^{*}+ \boldsymbol{B}_2^{*} + \boldsymbol{K}_{1}     \right) \right.\right. \nonumber \\
& \qquad\qquad\left.\left.    J   \left( \left. \sqrt{1-\gamma}X+\sqrt{\gamma}X^{G}_{1}+S^{(2)}_{1} \right| \sqrt{\gamma}X-\sqrt{1-\gamma}X_{1}^{G}+S^{(2)}_{2}, U  \right) \left( \boldsymbol{B}_1^{*}+ \boldsymbol{B}_2^{*} + \boldsymbol{K}_{1}     \right)  -\left(  \boldsymbol{B}_1^{*}+\boldsymbol{B}_2^{*} + \boldsymbol{K}_{1}     \right)\right)     \right\}, \label{eq:term3}
\end{align}
where (a) is due to KKT conditions \eqref{KKT_eq1} and \eqref{KKT_eq2} in Lemma \ref{lemma_KKT1}.
From data processing inequality in Lemma \ref{DP_FI},
\begin{align}
&J   \left(  \sqrt{1-\gamma}X+\sqrt{\gamma}X^{G}_{1}+S^{(2)}_{1} \left|   \sqrt{\gamma}X-\sqrt{1-\gamma}X_{1}^{G}+S^{(2)}_{2}, U \right. \right)\nonumber \\
& \preceq J   \left(  \sqrt{1-\gamma}X+\sqrt{\gamma}X^{G}_{1}+S^{(2)}_{1} \left|   \sqrt{\gamma}X-\sqrt{1-\gamma}X_{1}^{G}+S^{(2)}_{2}, X^{G}_{1},U,V \right. \right) \\
&= J   \left(  \sqrt{1-\gamma}X+S^{(2)}_{1} \left|   \sqrt{\gamma}X+S^{(2)}_{2},U,V \right. \right).
\end{align}
Thus, \eqref{eq:term1} can be upper bounded by
\begin{align}
&\tr \left\{ \boldsymbol{M}_1   \left(  \left( \boldsymbol{B}_1^{*}+ \boldsymbol{B}_2^{*} + \boldsymbol{K}_{1}     \right) J   \left(  \sqrt{1-\gamma}X+\sqrt{\gamma}X^{G}_{1}+S^{(2)}_{1} \left|   \sqrt{\gamma}X-\sqrt{1-\gamma}X_{1}^{G}+S^{(2)}_{2}, U \right. \right)    \left( \boldsymbol{B}_1^{*}+ \boldsymbol{B}_2^{*} + \boldsymbol{K}_{1}     \right) \right.\right. \nonumber \\
&\qquad \qquad\left.\left. -   \left( \boldsymbol{B}_2^{*} + \boldsymbol{K}_{1}     \right) J   \left(  \sqrt{1-\gamma}X+S^{(2)}_{1} \left|   \sqrt{\gamma}X+S^{(2)}_{2}, U,V \right. \right)    \left(  \boldsymbol{B}_2^{*} + \boldsymbol{K}_{1}     \right) - \boldsymbol{B}_1^{*}                                   \right)\right\}  \\
& \leq \tr \left\{ \boldsymbol{M}_1   \left(  \left( \boldsymbol{B}_1^{*}+ \boldsymbol{B}_2^{*} + \boldsymbol{K}_{1}     \right) J   \left(  \sqrt{1-\gamma}X+S^{(2)}_{1} \left|   \sqrt{\gamma}X+S^{(2)}_{2}, U, V \right. \right)    \left( \boldsymbol{B}_1^{*}+ \boldsymbol{B}_2^{*} + \boldsymbol{K}_{1}     \right) \right.\right. \nonumber \\
&\qquad \qquad\left.\left. -   \left( \boldsymbol{B}_2^{*} + \boldsymbol{K}_{1}     \right) J   \left(  \sqrt{1-\gamma}X+S^{(2)}_{1} \left|   \sqrt{\gamma}X+S^{(2)}_{2}, U,V \right. \right)    \left(  \boldsymbol{B}_2^{*} + \boldsymbol{K}_{1}     \right) - \boldsymbol{B}_1^{*}                                   \right)\right\}\\
&\overset{(a)}=0,\label{term1}
\end{align}
where (a) is due to KKT condition $\boldsymbol{B}_{1}^{*}\boldsymbol{M}_{1} =\boldsymbol{M}_{1}\boldsymbol{B}_{1}^{*}=0$ in Lemma \ref{lemma_KKT1}.

From data processing inequality in Lemma \ref{DP_FI},
\begin{align}
& J   \left(  \sqrt{1-\gamma}X+S^{(2)}_{1} \left|   \sqrt{\gamma}X+S^{(2)}_{2}, U,V \right. \right)\nonumber \\
&= J \left(      X^{(2)}_{+,\gamma}+\sqrt{1-\gamma}N_{1} + \sqrt{\gamma} N_{1}^{G}      \left|   X^{(2)}_{-,\gamma}+\sqrt{\gamma}N_{2} - \sqrt{1-\gamma} N_{2}^{G},       U,V    \right.  \right) \\
&\preceq J\left(      X^{(2)}_{+,\gamma}+\sqrt{1-\gamma}N_{1} + \sqrt{\gamma} N_{1}^{G}      \left|   X^{(2)}_{-,\gamma}+\sqrt{\gamma}N_{2} - \sqrt{1-\gamma} N_{2}^{G}, X^{(2)}_{+,\gamma}      U,V    \right.  \right) \\
& \overset{(a)} = J \left(       \sqrt{1-\gamma}N_{1} + \sqrt{\gamma} N_{1}^{G}       \right)\\
&=\boldsymbol{K}_{1}^{-1}.
\end{align}
where (a) is from the fact that $\sqrt{1-\gamma}N_{1} + \sqrt{\gamma} N_{1}^{G}$ is independent of $\sqrt{\gamma}N_{2} - \sqrt{1-\gamma} N_{2}^{G}$.  Thus, \eqref{eq:term2} can be upper bounded by
\begin{align}
& \tr \left\{     \boldsymbol{M}_2 \left(  \left(  \boldsymbol{B}_2^{*} + \boldsymbol{K}_{1}     \right) J   \left(  \sqrt{1-\gamma}X+S^{(2)}_{1} \left|   \sqrt{\gamma}X+S^{(2)}_{2}, U,V \right. \right)  \left(  \boldsymbol{B}_2^{*} + \boldsymbol{K}_{1}     \right)  -\left(  \boldsymbol{B}_2^{*} + \boldsymbol{K}_{1}     \right)\right)     \right\} \nonumber \\
& \geq \tr \left\{     \boldsymbol{M}_2 \left(  \left(  \boldsymbol{B}_2^{*} + \boldsymbol{K}_{1}     \right) \boldsymbol{K}_{1}^{-1}  \left(  \boldsymbol{B}_2^{*} + \boldsymbol{K}_{1}     \right)  -\left(  \boldsymbol{B}_2^{*} + \boldsymbol{K}_{1}     \right)\right)     \right\} \\
& = \tr \left\{     \boldsymbol{M}_2   \left(  \boldsymbol{B}_2^{*} + \boldsymbol{K}_{1}     \right) \boldsymbol{K}_{1}^{-1}  \boldsymbol{B}_2^{*}          \right\} \\
&\overset{(a)}=0,\label{term2}
\end{align}
where (a) is due to KKT condition $\boldsymbol{B}_{2}^{*}\boldsymbol{M}_{2} =0$ in Lemma \ref{lemma_KKT1}.

From data processing inequality in Lemma \ref{DP_FI} and Cram\'{e}r-Rao inequality of Lemma \ref{cri}, it can be shown that
\begin{align}
& J   \left( \left. \sqrt{1-\gamma}X+\sqrt{\gamma}X^{G}_{1}+S^{(2)}_{1} \right| \sqrt{\gamma}X-\sqrt{1-\gamma}X_{1}^{G}+S^{(2)}_{2}, U  \right)^{-1} \nonumber \\
& \preceq J   \left(  \sqrt{1-\gamma}X+\sqrt{\gamma}X^{G}_{1}+S^{(2)}_{1} \right)^{-1}\\
&= \cov \left(    \sqrt{1-\gamma}X+\sqrt{\gamma}X^{G}_{1}+ \sqrt{\gamma}X_{2}^{G}+\sqrt{1-\gamma}N_{1} + \sqrt{\gamma} N^{G}_{1}    \right)\\
&=(1-\gamma) \cov(X) + \gamma (\boldsymbol{B}_{1}^{*} + \boldsymbol{B}_{2}^{*} )+ \boldsymbol{K}_{1} \\
& \preceq (1-\gamma) \boldsymbol{S} + \gamma (\boldsymbol{B}_{1}^{*} + \boldsymbol{B}_{2}^{*} )+ \boldsymbol{K}_{1}.
\end{align}
Thus, \eqref{eq:term3} can be lower bounded by
\begin{align}
& \tr \left\{     \boldsymbol{M}_{3} \left(  \left( \boldsymbol{B}_1^{*}+ \boldsymbol{B}_2^{*} + \boldsymbol{K}_{1}     \right) \right.\right. \nonumber \\
& \qquad\qquad\left.\left.    J   \left( \left. \sqrt{1-\gamma}X+\sqrt{\gamma}X^{G}_{1}+S^{(2)}_{1} \right| \sqrt{\gamma}X-\sqrt{1-\gamma}X_{1}^{G}+S^{(2)}_{2}, U  \right) \left( \boldsymbol{B}_1^{*}+ \boldsymbol{B}_2^{*} + \boldsymbol{K}_{1}     \right)  -\left(  \boldsymbol{B}_1^{*}+\boldsymbol{B}_2^{*} + \boldsymbol{K}_{1}     \right)\right)     \right\} \nonumber \\
& \geq \tr \left\{    \boldsymbol{M}_{3}  \left(  \left( \boldsymbol{B}_1^{*}+ \boldsymbol{B}_2^{*} + \boldsymbol{K}_{1}     \right) \left(   (1-\gamma) \boldsymbol{S} + \gamma (\boldsymbol{B}_{1}^{*} + \boldsymbol{B}_{2}^{*} )+ \boldsymbol{K}_{1} \right)^{-1} \left( \boldsymbol{B}_1^{*}+ \boldsymbol{B}_2^{*} + \boldsymbol{K}_{1}     \right)  -\left(  \boldsymbol{B}_1^{*}+\boldsymbol{B}_2^{*} + \boldsymbol{K}_{1}     \right)\right)     \right\} \\
&= -\tr\left\{  \boldsymbol{M}_{3} (1-\gamma)\left( \boldsymbol{B}_1^{*}+ \boldsymbol{B}_2^{*} + \boldsymbol{K}_{1}     \right) \left(   (1-\gamma) \boldsymbol{S} + \gamma (\boldsymbol{B}_{1}^{*} + \boldsymbol{B}_{2}^{*} )+ \boldsymbol{K}_{1} \right)^{-1}\left( \boldsymbol{S} - \boldsymbol{B}^{*}_{1} - \boldsymbol{B}^{*}_{2}\right)           \right\}\\
&\overset{(a)}=0,\label{term3}
\end{align}
where (a) is due to KKT condition $\left( \boldsymbol{S} - \boldsymbol{B}^{*}_{1} - \boldsymbol{B}^{*}_{2}\right)\boldsymbol{M}_{3} =0$ in Lemma \ref{lemma_KKT1}.
Combining \eqref{term1}, \eqref{term2} and \eqref{term3}, this completes the perturbation proof of $dg(\gamma)/d\gamma \geq 0$, and so the extremal inequality \eqref{eq:m3} in Theorem \ref{mainthm3}.

\section{Rate-Distortion-Equivocation Function of the Vector Gaussian Secure Source Coding}\label{sec5}

\subsection{Problem Statement}

The vector Gaussian secure source coding problem setup consists of one encoder, one legitimate decoder and one eavesdropper decoder. Let $\{{X}, {Y}, {Z}\}$ be a tuple of random vectors, which is drawn from a jointly vector Gaussian distribution. The encoder, the legitimate decoder and the eavesdropper decoder observe ${X}$, ${Y}$ and ${Z}$, respectively.

The vector Gaussian source $\{{X}, {Y}, {Z}\}$ can be written as
\begin{align}
{Y} =  {X} + {N}_{Y}, \label{eq:source1}\\
{Z} =  {X} + {N}_{Z}, \label{eq:source2}
\end{align}
where ${X}$ is a $p\times1$-dimensional Gaussian random vector with mean zero and covariance $\boldsymbol{K} \succ0$, each ${N}_{Y}$ is a $p\times1$-dimensional Gaussian random vector with mean zero and covariance $\boldsymbol{K}_{Y} \succ 0 $, and ${N}_{Z}$ is a $p\times1$-dimensional Gaussian random vector with mean zero and covariance $\boldsymbol{K}_{Z} \succ 0 $, respectively. We shall point out that $({N}_{Y}, {N}_{Z})$ and ${X}$ are independent from expressions \eqref{eq:source1} and \eqref{eq:source2}. However, no additional independence relationship is imposed between ${N}_{Y}(t)$ and ${N}_{Z}(t)$.

The encoder wants to convey an $n$-length source sequence ${X}^{n} \triangleq ( X_{1}, \ldots, {X}_{n} )$ to the legitimate decoder within a distortion constraint, and meanwhile the eavesdropper decoder is kept ignorant of source ${X}^{n}$, which is measured by equivocation.

A $(2^{nR},n)$ code of rate $R$ consists of
\begin{itemize}
 \item an encoding function $\phi: \mathcal{R}^{p \times n} \mapsto \mathcal{M}^{n} = \{ 1, \ldots, 2^{nR}\}$ that finds a codeword $m(x^{n})$ to each n-length source sequence $x^{n}$, and sends it to both decoders,
  \item an legitimate decoding function $\psi: \mathcal{R}^{p \times n} \times  \mathcal{M}^{n} \mapsto \mathcal{R}^{p \times n}$ that assigns an estimate $\hat{x}^{n} (m, y^{n})$ to each received codeword $m$ and the side information $y^{n}$.
\end{itemize}
The distortion of the reconstructed $\hat{X}$ at the legitimate is measured by the mean square error (MSE) matrix $\boldsymbol{D}$,
and the privacy leakage about the source ${{X}}$ at the eavesdropper is measured by the equivocation rate $R_{e}$ as in \cite{Wyner75}.
A rate-distortion-equivocation tuple $(R, \boldsymbol{D}, R_{e})$ is said to be achievable if there exists a sequence of $(2^{nR},n)$ code such that
\begin{align}
\frac{1}{n} \cov \left( {X}^n | {Y}^{n}, M  \right) \preceq \boldsymbol{D}, \\
\frac{1}{n} H({X}^{n} | {Z}^{n}, M) \geq R_{e}.
\end{align}

For a fixed ${\boldsymbol{D}}$, the rate-equivocation pair $(R, R_{e})$ is included in region $\mathcal{R}({\boldsymbol{D}})$. Since region $\mathcal{R}({\boldsymbol{D}})$ is convex, to characterize the rate-distortion-equivocation function for the vector Gaussian model, we can alternatively consider the following $\mu$-difference problem
\begin{equation}
\inf_{(R, R_{e} )\in \mathcal{R}({\boldsymbol{D}})} \mu R-R_{e},
\end{equation}
for any $\mu \geq 0$.

\subsection{The Extremal Inequality}

Villard and Piantanida studied the general setting of secure source coding problem in \cite{Villard13}. By performing rate splitting and Wyner-Ziv coding to exploit the side information at the legitimate, an achievable rate-distortion-equivocation tradeoff region can be characterized by the following matrix optimization problem,
\begin{align}
\inf_{(R, R_{e} )\in \mathcal{R}({\boldsymbol{D}})} \mu R-R_{e} \leq R^{*}({\boldsymbol{D}}, \mu),
\end{align}
where
\begin{align} \label{eqn:opt}
& R^{*}({\boldsymbol{D}}, \mu) \nonumber \\
\triangleq &\min_{\boldsymbol{B}_{1},\boldsymbol{B}_{2}}\frac{\mu+1}{2} \log \left| (2\pi e)\left( \boldsymbol{K}^{-1} + \boldsymbol{K}^{-1}_{Y} + \boldsymbol{B}_{1}+\boldsymbol{B}_{2}\right) \right|-\frac{1}{2} \log \left|(2 \pi e)\left(\boldsymbol{K}^{-1} + \boldsymbol{K}^{-1}_{Y} +\boldsymbol{B}_{2}\right) \right| \nonumber \\
&\qquad\quad +\frac{1}{2} \log \left|(2 \pi e)\left(\boldsymbol{K}^{-1} + \boldsymbol{K}^{-1}_{Z} +\boldsymbol{B}_{2}\right) \right| - \frac{\mu}{2} \log \left| (2 \pi e) \left(\boldsymbol{K}^{-1} + \boldsymbol{K}^{-1}_{Y} \right) \right| \nonumber \\
&\;\text{subject to}\quad \boldsymbol{B}_{1} \succeq {0}, \; \boldsymbol{B}_{2} \succeq {0},  \nonumber \\
& \qquad \qquad \quad \;\boldsymbol{B}_{1}+\boldsymbol{B}_{2} \succeq \boldsymbol{D}^{-1} -\boldsymbol{K}^{-1} - \boldsymbol{K}^{-1}_{Y}.
\end{align}

To establish an extremal inequality and characterize the rate-distortion-equivocation function of the vector Gaussian secure source coding problem, we need further show
\begin{align}
\inf_{(R, R_{e} )\in \mathcal{R}({\boldsymbol{D}})} \mu R-R_{e} \geq R^{*}({\boldsymbol{D}}, \mu).
\end{align}

In \cite{EU13}, Ekrem and Ulukus studied the vector Gaussian secure source coding problem. and partially characterize the rate-distortion-equivocation function in the case of $\mu=1$. However, it seems a difficult task to generalize their source enhancement argument beyond the case of $\mu=1$. As pointed out in \cite{XC21}, it is also unclear how to generalize the method of factorization in \cite{GN14} to handle the non-degraded source. In this section, we proof the extremal inequality based on a novel monotone path construction, and fully characterize the rate-distortion-equivocation function for arbitrary positive $\mu$.

Let $(\boldsymbol{B}_{1}^{*},\boldsymbol{B}_{2}^{*})$ be one minimizer of the optimization problem $R^{*}({\boldsymbol{D}}, \mu)$. The necessary Karush-Kuhn-Tucker (KKT) conditions are given in the following lemma, whose proof is omitted because it is a standard evaluation on Lagrange function of $R^{*}({\boldsymbol{D}}, \mu)$.
\begin{lemma} \label{lemma_KKT} The minimizer $(\boldsymbol{B}_{1}^{*},\boldsymbol{B}_{2}^{*})$ of $R^{*}({\boldsymbol{D}}, \mu)$ need to satisfy
\begin{align}
(\mu+1) \left(\boldsymbol{K}^{-1} + \boldsymbol{K}_{Y}^{-1} + \boldsymbol{B}_{1}^{*}+ \boldsymbol{B}_{2}^{*} \right)^{-1} &= \boldsymbol{M}_{1} + \boldsymbol{M}_{3}, \label{eq:KKT1}\\
\left(\boldsymbol{K}^{-1} + \boldsymbol{K}_{Y}^{-1} + \boldsymbol{B}_{2}^{*} \right)^{-1} + \boldsymbol{M}_{2} &= \left(\boldsymbol{K}^{-1} + \boldsymbol{K}_{Z}^{-1} + \boldsymbol{B}_{2}^{*} \right)^{-1} + \boldsymbol{M}_{1}, \label{eq:KKT2}
\end{align}
for some positive semi-definite matrices $\boldsymbol{M}_{1}, \boldsymbol{M}_{2}, \boldsymbol{M}_{3} \succeq0$ such that
\begin{align}
\boldsymbol{B}_{1}^{*} \boldsymbol{M}_{1} &=0, \label{eq:KKT3}\\
\boldsymbol{B}_{2}^{*} \boldsymbol{M}_{2}&=0, \label{eq:KKT4}\\
\left(\boldsymbol{K}^{-1} + \boldsymbol{K}^{-1}_{Y}+\boldsymbol{B}_{1}+\boldsymbol{B}_{2} - \boldsymbol{D}^{-1}\right) \boldsymbol{M}_3&=0 \label{eq:KKT5}.
\end{align}
\end{lemma}

Now starting from the single-letter expressions in \cite[Theroem 3]{Villard13}, the $\mu$-difference of $\mu R- R_{e}$ for any rate-equivocation pair $(R, R_{e}) \in \mathcal{R}({\boldsymbol{D}})$ should be lower bounded by
\begin{align}
& \mu R- R_{e} \nonumber \\
\geq \;& \mu I({X}; V | {Y}) - h({X} | V,{Y})- I({X}; {Y}| U)+ I({X};{Z}|U) \\
=\; & (\mu+1) I({X}; V| {Y}) - h({X} | {Y})- I({X}; {Y}| U) + I({X};{Z}|U) \\
=\; & -(\mu+1)h(X|Y,V)+h(X|Y,U)-h(X|Z,U)+\mu h(X|Y)\\
=\; & -(\mu+1)h(X|Y,V)+h(X|Y,U)-h(X|Z,U)-\frac{\mu}{2}\log \left|   (2 \pi e) \left(\boldsymbol{K}^{-1} +\boldsymbol{K}_{Y}^{-1}\right)   \right|. \label{eqn:lb}
\end{align}

By comparing \eqref{eqn:lb} with optimization problem $R^{*}({\boldsymbol{D}}, \mu)$ in \eqref{eqn:opt}, it can be shown that to prove to converse part, it is sufficient to prove the following extremal inequality.

\begin{theorem} \label{ext_thm}
There exist two positive semi-definite matrices $\boldsymbol{B}_{1}^{*}$ and $\boldsymbol{B}_{2}^{*}$, which satisfy KKT conditions \eqref{eq:KKT1}-\eqref{eq:KKT5} in lemma \ref{lemma_KKT} to minimize optimization problem $R^{*}({\boldsymbol{D}}, \mu)$, then for some real number $\mu \geq 0$, we have
\begin{align}
&-(\mu+1)h(X|Y,V)+h(X|Y,U)-h(X|Z,U) \nonumber \\
\geq \;& \frac{\mu+1}{2} \log \left| (2\pi e)\left( \boldsymbol{K}^{-1} + \boldsymbol{K}^{-1}_{Y} + \boldsymbol{B}^{*}_{1}+\boldsymbol{B}^{*}_{2}\right) \right|-\frac{1}{2} \log \left|(2 \pi e)\left(\boldsymbol{K}^{-1} + \boldsymbol{K}^{-1}_{Y} +\boldsymbol{B}^{*}_{2}\right) \right|  \nonumber \\
&+\frac{1}{2} \log \left|(2 \pi e)\left(\boldsymbol{K}^{-1} + \boldsymbol{K}^{-1}_{Z} +\boldsymbol{B}^{*}_{2}\right) \right|, \label{eq:exinq}
\end{align}
for any $(U,V)$ such that $U \rightarrow V \rightarrow {X} \rightarrow ({Y}, {Z})$ forms a Markov chain and $\cov ({X} | {Y},V) \preceq  {\boldsymbol{D}}$.
\end{theorem}

\subsection {Proof of Theorem \ref{ext_thm}}
\subsubsection{Monotone Path Construction}
For the sake of simplifying notations, we firstly denote by
\begin{align}
\boldsymbol{\Delta}^{-1}_{1} &= \boldsymbol{K}^{-1} + \boldsymbol{B}_{1}^{*} + \boldsymbol{B}_{2}^{*}, \\
\boldsymbol{\Delta}^{-1}_{2} &= \boldsymbol{K}^{-1} + \boldsymbol{B}_{2}^{*}.
\end{align}
We consider the covariance preserved transform $\left\{X^{(1)}_{+,\gamma}, X^{(1)}_{-,\gamma}, X^{(2)}_{+,\gamma}, X^{(2)}_{-,\gamma} \right\}$ as
\begin{align}
X^{(1)}_{+, \gamma} &= \sqrt{1-\gamma}X + \sqrt{\gamma}X^{G}_{1}, \\
X^{(1)}_{-, \gamma} &= \sqrt{\gamma}X - \sqrt{1-\gamma}X^{G}_{1},\\
X^{(2)}_{+, \gamma} &= X^{(1)}_{+, \gamma} +  \sqrt{\gamma}X^{G}_{2},\\
X^{(2)}_{-, \gamma} &= X^{(2)}_{-, \gamma} -\sqrt{1-\gamma}X^{G}_{2},
\end{align}
where $X^{G}_{1}$ and $X^{G}_{2}$ are mutually independent random vectors, which follow Gaussian distributions $\mathcal{N}(0, \boldsymbol{\Delta}_{1})$ and $\mathcal{N}(0, \boldsymbol{\Delta}_{2}-\boldsymbol{\Delta}_{1})$, separately. For any $\gamma \in (0,1)$, the perturbation function $g(\gamma)$ is defined as
\begin{align}
g(\gamma) &= -(\mu+1)h\left(\left.X^{(1)}_{+, \gamma}, X^{(1)}_{-, \gamma}+\sqrt{\gamma}N_{Y} - \sqrt{1-\gamma}N_{Y}^{G} \right| V \right)\nonumber \\
& + h\left(\left.X^{(2)}_{+, \gamma}+\sqrt{1-\gamma}N_{Z}+ \sqrt{\gamma}N_{Z}^{G}, X^{(2)}_{-, \gamma}+\sqrt{\gamma}N_{Y} - \sqrt{1-\gamma}N_{Y}^{G} \right| U \right),
\end{align}
where $N_{Y}^{G}$, $N_{Z}^{G}$ are independent Gaussian random vectors with the same distributions of $N_{Y}$, $N_{Z}$.

When $\gamma=0$, we have
\begin{align}
g(0)=-(\mu+1)h\left( X |V      \right) - (\mu+1)h(X_{1}^{G} + N_{Y}^{G})+h(Z|U)+h(X_{1}^{G}+X_{2}^{G} + N_{Y}^{G})
\end{align}
When $\gamma=1$, we have
\begin{align}
g(1)=-(\mu+1)h(X^{G})-(\mu+1)h(Y|V)+h(X_{1}^{G}+X_{2}^{G}+N_{Z}^{G})+h(Y|U).
\end{align}
Therefore,
\begin{align}
&g(0)-g(1) \nonumber \\
&=-(\mu+1)\left((h(X|V)-h(Y|V)\right)+h(Z|U)-h(Y|U)\nonumber \\
& \quad -\frac{\mu+1}{2} \log \frac {\left|  \boldsymbol{\Delta}_{1} + \boldsymbol{K}_{Y}   \right|}{\left|  \boldsymbol{\Delta}_{1} \right|} +\frac{1}{2} \log \frac {\left|  \boldsymbol{\Delta}_{2} + \boldsymbol{K}_{Y}   \right|}{\left|  \boldsymbol{\Delta}_{2} +\boldsymbol{K}_{Z} \right|} \\
&=-(\mu+1) h(X|Y,V)+h(X|Y,U)-h(X|Z,U)\nonumber \\
& \quad -\frac{\mu+1}{2} \log \left| (2 \pi e) \left(    \boldsymbol{\Delta}^{-1}_{1} + \boldsymbol{K}^{-1}_{Y}    \right)    \right| + \frac{1}{2} \log \left| (2 \pi e) \left(    \boldsymbol{\Delta}^{-1}_{2} + \boldsymbol{K}^{-1}_{Y}    \right)    \right|- \frac{1}{2} \log \left| (2 \pi e) \left(    \boldsymbol{\Delta}^{-1}_{2} + \boldsymbol{K}^{-1}_{Z}    \right)    \right|.\label{eq:qq2}
\end{align}
Comparing \eqref{eq:exinq} with \eqref{eq:qq2}, it can be concluded that it needs to prove $g(\gamma)$ is a monotonically decreasing function on $\gamma$, \emph{i.e.},
\begin{equation}
\frac{d}{d\gamma}g(\gamma) \leq 0, \quad \gamma \in (0,1).
\end{equation}

\subsubsection{Derivative Evaluation}
Using the same method of calculating the derivative from \eqref{eqn:subto_2} to \eqref{eq:tem1}, we can obtain
\begin{align}
& \frac{d}{d\gamma} h\left(\left.X^{(1)}_{+, \gamma}, X^{(1)}_{-, \gamma}+\sqrt{\gamma}N_{Y} - \sqrt{1-\gamma}N_{Y}^{G} \right| V \right)\nonumber \\
& = \frac{1}{2(1-\gamma)} \tr \left\{  \left(   \boldsymbol{\Delta}_{1}^{-1} - \left( \boldsymbol{\Delta}_{1} + \boldsymbol{K}_{Y}  \right)^{-1}        \right) \left(       \boldsymbol{\Delta}_{1} J \left(  \left. X^{(1)}_{+, \gamma}  \right| X^{(1)}_{-, \gamma}+\sqrt{\gamma}N_{Y} - \sqrt{1-\gamma}N_{Y}^{G}, V  \right)   \boldsymbol{\Delta}_{1} - \boldsymbol{\Delta}_{1}   \right)                          \right\} \\
& \overset{(a)}= \frac{1}{2(1-\gamma)} \tr \left\{   \left( \boldsymbol{\Delta}^{-1}_{1} + \boldsymbol{K}^{-1}_{Y}  \right)^{-1}\left(  J \left(  \left. X^{(1)}_{+, \gamma}  \right| X^{(1)}_{-, \gamma}+\sqrt{\gamma}N_{Y} - \sqrt{1-\gamma}N_{Y}^{G}, V  \right) - \boldsymbol{\Delta}^{-1}_{1}  \right)           \right\},
\end{align}
where (a) is due Woodbury matrix inversion identity.

We can similarly obtain
\begin{align}
&\frac{d}{d\gamma} h\left(\left.X^{(2)}_{+, \gamma}+\sqrt{1-\gamma}N_{Z}+ \sqrt{\gamma}N_{Z}^{G}, X^{(2)}_{-, \gamma}+\sqrt{\gamma}N_{Y} - \sqrt{1-\gamma}N_{Y}^{G} \right| V \right)\nonumber \\
&=\frac{1}{2(1-\gamma)} \tr \left\{   \left(      \left( \boldsymbol{\Delta}_{2} + \boldsymbol{K}_{Z}  \right)^{-1} - \left( \boldsymbol{\Delta}_{2} + \boldsymbol{K}_{Y}  \right)^{-1}        \right)\left(\left( \boldsymbol{\Delta}_{2} + \boldsymbol{K}_{Z}  \right)\right.\right.\nonumber \\
& \qquad  \left.\left.J\left( \left.X^{(2)}_{+, \gamma}+\sqrt{1-\gamma}N_{Z}+ \sqrt{\gamma}N_{Z}^{G} \right| X^{(2)}_{-, \gamma}+\sqrt{\gamma}N_{Y} - \sqrt{1-\gamma}N_{Y}^{G},U \right)\left( \boldsymbol{\Delta}_{2} + \boldsymbol{K}_{Z}  \right)-\left( \boldsymbol{\Delta}_{2} + \boldsymbol{K}_{Z}  \right) \right)        \right\} \\
&=\frac{1}{2(1-\gamma)} \tr \left\{   \left(      \left( \boldsymbol{\Delta}^{-1}_{2} + \boldsymbol{K}^{-1}_{Y}  \right)^{-1} - \left( \boldsymbol{\Delta}^{-1}_{2} + \boldsymbol{K}^{-1}_{Z}  \right)^{-1}        \right)\left(\left( \boldsymbol{I} + \boldsymbol{\Delta}^{-1}_{2}\boldsymbol{K}_{Z}  \right)\right.\right.\nonumber \\
& \qquad   \left.\left.J\left( \left.X^{(2)}_{+, \gamma}+\sqrt{1-\gamma}N_{Z}+ \sqrt{\gamma}N_{Z}^{G} \right|X^{(2)}_{-, \gamma}+\sqrt{\gamma}N_{Y} - \sqrt{1-\gamma}N_{Y}^{G}, U \right)\left( \boldsymbol{I} + \boldsymbol{K}_{Z}\boldsymbol{\Delta}^{-1}_{2}  \right)-\boldsymbol{\Delta}_{2}^{-1}\left( \boldsymbol{\Delta}_{2} + \boldsymbol{K}_{Z}  \right)\boldsymbol{\Delta}_{2}^{-1} \right)        \right\}.
\end{align}
So, the derivative of $g(\gamma)$ can be written as
\begin{align}
& 2(1-\gamma) \frac{d}{d\gamma}g(\gamma)\nonumber \\
&= \tr \left\{  -(\mu+1) \left( \boldsymbol{\Delta}^{-1}_{1} + \boldsymbol{K}^{-1}_{Y}  \right)^{-1}\left(  J \left(  \left. X^{(1)}_{+, \gamma}  \right| X^{(1)}_{-, \gamma}+\sqrt{\gamma}N_{Y} - \sqrt{1-\gamma}N_{Y}^{G}, V  \right) - \boldsymbol{\Delta}^{-1}_{1}  \right)           \right\} \\
& \quad +\tr \left\{   \left(      \left( \boldsymbol{\Delta}^{-1}_{2} + \boldsymbol{K}^{-1}_{Y}  \right)^{-1} - \left( \boldsymbol{\Delta}^{-1}_{2} + \boldsymbol{K}^{-1}_{Z}  \right)^{-1}        \right)\left(\left( \boldsymbol{I} + \boldsymbol{\Delta}^{-1}_{2}\boldsymbol{K}_{Z}  \right)\right.\right.\nonumber \\
& \qquad   \left.\left.J\left( \left.X^{(2)}_{+, \gamma}+\sqrt{1-\gamma}N_{Z}+ \sqrt{\gamma}N_{Z}^{G} \right|X^{(2)}_{-, \gamma}+\sqrt{\gamma}N_{Y} - \sqrt{1-\gamma}N_{Y}^{G}, U \right)\left( \boldsymbol{I} + \boldsymbol{K}_{Z}\boldsymbol{\Delta}^{-1}_{2}  \right)-\boldsymbol{\Delta}_{2}^{-1}\left( \boldsymbol{\Delta}_{2} + \boldsymbol{K}_{Z}  \right)\boldsymbol{\Delta}_{2}^{-1} \right)        \right\} \\
&=\tr \left\{   \boldsymbol{M}_{1}\left(\left( \boldsymbol{I} + \boldsymbol{\Delta}^{-1}_{2}\boldsymbol{K}_{Z}  \right) J\left( \left.X^{(2)}_{+, \gamma}+\sqrt{1-\gamma}N_{Z}+ \sqrt{\gamma}N_{Z}^{G} \right|X^{(2)}_{-, \gamma}+\sqrt{\gamma}N_{Y} - \sqrt{1-\gamma}N_{Y}^{G}, U \right)\left( \boldsymbol{I} + \boldsymbol{K}_{Z}\boldsymbol{\Delta}^{-1}_{2}  \right)\right.\right.\nonumber \\
& \hspace{0.7in} \left.\left. -  J \left(  \left. X^{(1)}_{+, \gamma}  \right| X^{(1)}_{-, \gamma}+\sqrt{\gamma}N_{Y} - \sqrt{1-\gamma}N_{Y}^{G}, V  \right) -\boldsymbol{\Delta}_{2}^{-1}\boldsymbol{K}_{Z}\boldsymbol{\Delta}_{2}^{-1}- \boldsymbol{\Delta}_{2}^{-1} + \boldsymbol{\Delta}_{1}^{-1}\right)        \right\} \label{eq:te1}\\
& \quad - \tr \left\{   \boldsymbol{M}_{2}\left(\left( \boldsymbol{I} + \boldsymbol{\Delta}^{-1}_{2}\boldsymbol{K}_{Z}  \right)J\left( \left.X^{(2)}_{+, \gamma}+\sqrt{1-\gamma}N_{Z}+ \sqrt{\gamma}N_{Z}^{G} \right|X^{(2)}_{-, \gamma}+\sqrt{\gamma}N_{Y} - \sqrt{1-\gamma}N_{Y}^{G}, U \right)\left( \boldsymbol{I} + \boldsymbol{K}_{Z}\boldsymbol{\Delta}^{-1}_{2}  \right)\right.\right.\nonumber \\
& \hspace{0.5in} \qquad   \left.\left.-\boldsymbol{\Delta}_{2}^{-1}\left( \boldsymbol{\Delta}_{2} + \boldsymbol{K}_{Z}  \right)\boldsymbol{\Delta}_{2}^{-1} \right)        \right\} \label{eq:te2}\\
& \quad -\tr \left\{   \boldsymbol{M}_{3}\left(  J \left(  \left. X^{(1)}_{+, \gamma}  \right| X^{(1)}_{-, \gamma}+\sqrt{\gamma}N_{Y} - \sqrt{1-\gamma}N_{Y}^{G}, V  \right) - \boldsymbol{\Delta}^{-1}_{1}  \right)           \right\} \label{eq:te3}.
\end{align}

\subsubsection{Upper Bounds}
From data processing inequality in Lemma \ref{DP_FI}, and Markov chain $U \rightarrow V \rightarrow {X}$, we have
\begin{align}
& J\left( \left.X^{(2)}_{+, \gamma}+\sqrt{1-\gamma}N_{Z}+ \sqrt{\gamma}N_{Z}^{G} \right|X^{(2)}_{-, \gamma}+\sqrt{\gamma}N_{Y} - \sqrt{1-\gamma}N_{Y}^{G}, U \right) \nonumber \\
&\preceq J \left(     \left.X^{(2)}_{+, \gamma}+\sqrt{1-\gamma}N_{Z}+ \sqrt{\gamma}N_{Z}^{G} \right|X^{(2)}_{-, \gamma}+\sqrt{\gamma}N_{Y} - \sqrt{1-\gamma}N_{Y}^{G}, X_{2}^{G}, V       \right)\\
& = J \left(     \left.X^{(1)}_{+, \gamma}+\sqrt{1-\gamma}N_{Z}+ \sqrt{\gamma}N_{Z}^{G} \right|X^{(1)}_{-, \gamma}+\sqrt{\gamma}N_{Y} - \sqrt{1-\gamma}N_{Y}^{G}, V       \right). \label{eq:221}
\end{align}
Since $\sqrt{1-\gamma}N_{Z}+ \sqrt{\gamma}N_{Z}^{G}$ is independent of $X^{(1)}_{-, \gamma}$ and $\sqrt{\gamma}N_{Y} - \sqrt{1-\gamma}N_{Y}^{G}$, we can apply Lemma \ref{fi_inq} in Appendix \ref{app_lea2}. It is shown that
\begin{align}
&\left( \boldsymbol{I} + \boldsymbol{\Delta}^{-1}_{2}\boldsymbol{K}_{Z}  \right) J \left(     \left.X^{(1)}_{+, \gamma}+\sqrt{1-\gamma}N_{Z}+ \sqrt{\gamma}N_{Z}^{G} \right|X^{(1)}_{-, \gamma}+\sqrt{\gamma}N_{Y} - \sqrt{1-\gamma}N_{Y}^{G}, V       \right)\left( \boldsymbol{I} + \boldsymbol{K}_{Z}\boldsymbol{\Delta}^{-1}_{2}  \right) \nonumber \\
& \preceq J \left(     \left.X^{(1)}_{+, \gamma}\right|X^{(1)}_{-, \gamma}+\sqrt{\gamma}N_{Y} - \sqrt{1-\gamma}N_{Y}^{G}, V       \right)+ \boldsymbol{\Delta}^{-1}_{2}\boldsymbol{K}_{Z}\boldsymbol{\Delta}^{-1}_{2}. \label{eq:222}
\end{align}
By combining \eqref{eq:221} and \eqref{eq:222}, it shows that \eqref{eq:te1} can be bounded by
\begin{align}
&\tr \left\{   \boldsymbol{M}_{1}\left(\left( \boldsymbol{I} + \boldsymbol{\Delta}^{-1}_{2}\boldsymbol{K}_{Z}  \right) J\left( \left.X^{(2)}_{+, \gamma}+\sqrt{1-\gamma}N_{Z}+ \sqrt{\gamma}N_{Z}^{G} \right|X^{(2)}_{-, \gamma}+\sqrt{\gamma}N_{Y} - \sqrt{1-\gamma}N_{Y}^{G}, U \right)\left( \boldsymbol{I} + \boldsymbol{K}_{Z}\boldsymbol{\Delta}^{-1}_{2}  \right)\right.\right.\nonumber \\
& \hspace{0.7in} \left.\left. -  J \left(  \left. X^{(1)}_{+, \gamma}  \right| X^{(1)}_{-, \gamma}+\sqrt{\gamma}N_{Y} - \sqrt{1-\gamma}N_{Y}^{G}, V  \right) -\boldsymbol{\Delta}_{2}^{-1}\boldsymbol{K}_{Z}\boldsymbol{\Delta}_{2}^{-1}- \boldsymbol{\Delta}_{2}^{-1} + \boldsymbol{\Delta}_{1}^{-1}\right)        \right\}\nonumber \\
& \leq \tr \left\{   \boldsymbol{M}_{1}    \left(      \boldsymbol{\Delta}_{1}^{-1}- \boldsymbol{\Delta}_{2}^{-1}    \right)          \right\}= \tr \left\{  \boldsymbol{M}_{1} \boldsymbol{B}_{1}^{*}\right\}=0. \label{eq:tter1}
\end{align}

Again from data processing inequality in Lemma \ref{DP_FI}, we have
\begin{align}
& J\left( \left.X^{(2)}_{+, \gamma}+\sqrt{1-\gamma}N_{Z}+ \sqrt{\gamma}N_{Z}^{G} \right|X^{(2)}_{-, \gamma}+\sqrt{\gamma}N_{Y} - \sqrt{1-\gamma}N_{Y}^{G}, U \right)\nonumber \\
& \succeq J\left( \left.X^{(2)}_{+, \gamma}+\sqrt{1-\gamma}N_{Z}+ \sqrt{\gamma}N_{Z}^{G} \right.  \right)\\
&=\left((1-\gamma)\boldsymbol{K} + \gamma \boldsymbol{\Delta}_{2} + \boldsymbol{K}_{Z} \right)^{-1} \label{eq:asd}
\end{align}
Substituting \eqref{eq:asd} into \eqref{eq:te2}, it can be shown
\begin{align}
& \tr \left\{   \boldsymbol{M}_{2}\left(\left( \boldsymbol{I} + \boldsymbol{\Delta}^{-1}_{2}\boldsymbol{K}_{Z}  \right)J\left( \left.X^{(2)}_{+, \gamma}+\sqrt{1-\gamma}N_{Z}+ \sqrt{\gamma}N_{Z}^{G} \right|X^{(2)}_{-, \gamma}+\sqrt{\gamma}N_{Y} - \sqrt{1-\gamma}N_{Y}^{G}, U \right)\left( \boldsymbol{I} + \boldsymbol{K}_{Z}\boldsymbol{\Delta}^{-1}_{2}  \right)\right.\right.\nonumber \\
& \hspace{0.5in} \qquad   \left.\left.-\boldsymbol{\Delta}_{2}^{-1}\left( \boldsymbol{\Delta}_{2} + \boldsymbol{K}_{Z}  \right)\boldsymbol{\Delta}_{2}^{-1} \right)        \right\} \nonumber \\
& \geq \tr \left\{   \boldsymbol{M}_{2}\left(\left( \boldsymbol{I} + \boldsymbol{\Delta}^{-1}_{2}\boldsymbol{K}_{Z}  \right) \left((1-\gamma)\boldsymbol{K} + \gamma \boldsymbol{\Delta}_{2} + \boldsymbol{K}_{Z} \right)^{-1}\left( \boldsymbol{I} + \boldsymbol{K}_{Z}\boldsymbol{\Delta}^{-1}_{2}  \right)-\boldsymbol{\Delta}_{2}^{-1}\left( \boldsymbol{\Delta}_{2} + \boldsymbol{K}_{Z}  \right)\boldsymbol{\Delta}_{2}^{-1} \right)        \right\} \\
&=\tr \left\{   \boldsymbol{M}_{2}\left( \boldsymbol{I} + \boldsymbol{\Delta}^{-1}_{2}\boldsymbol{K}_{Z}  \right) \left((1-\gamma)\boldsymbol{K} + \gamma \boldsymbol{\Delta}_{2} + \boldsymbol{K}_{Z} \right)^{-1}
 \left( \boldsymbol{\Delta}_{2} + \boldsymbol{K}_{Z} - \left( (1-\gamma)\boldsymbol{K} + \gamma \boldsymbol{\Delta}_{2} + \boldsymbol{K}_{Z}\right)       \right) \boldsymbol{\Delta}_{2}^{-1}        \right\}\\
 &=\tr \left\{  (1-\gamma) \boldsymbol{M}_{2}\left( \boldsymbol{I} + \boldsymbol{\Delta}^{-1}_{2}\boldsymbol{K}_{Z}  \right) \left((1-\gamma)\boldsymbol{K} + \gamma \boldsymbol{\Delta}_{2} + \boldsymbol{K}_{Z} \right)^{-1}\boldsymbol{K}^{-1}
 \left(  \boldsymbol{K}^{-1} - \boldsymbol{\Delta}_{2}^{-1}    \right)        \right\}\\
 &=\tr \left\{  (1-\gamma) \boldsymbol{M}_{2}\left( \boldsymbol{I} + \boldsymbol{\Delta}^{-1}_{2}\boldsymbol{K}_{Z}  \right) \left((1-\gamma)\boldsymbol{K} + \gamma \boldsymbol{\Delta}_{2} + \boldsymbol{K}_{Z} \right)^{-1}\boldsymbol{K}^{-1}
 \boldsymbol{B}^{*}_{2}  \right\}\\
 &\overset{(a)}=0,\label{eq:tter2}
\end{align}
where (a) is due to KKT condition $\boldsymbol{B}_{2}^{*} \boldsymbol{M}_{2}=0$ in \eqref{eq:KKT4}.

Now using a similar argument as \eqref{eqn:su_1}-\eqref{eqn:su_2}, we show that
\begin{align}
& J \left(  \left. X^{(1)}_{+, \gamma}  \right| X^{(1)}_{-, \gamma}+\sqrt{\gamma}N_{Y} - \sqrt{1-\gamma}N_{Y}^{G}, V  \right) -\boldsymbol{\Delta}_{1}^{-1} \nonumber \\
&\overset{(a)}=(1-\gamma)\left(  \boldsymbol{\Delta}_{1} + (1-\gamma) \boldsymbol{K}_{Y}  \right)^{-1}\left(  \boldsymbol{\Delta}_{1} +  \boldsymbol{K}_{Y} \right)J (X +W|V) \left(  \boldsymbol{\Delta}_{1} +  \boldsymbol{K}_{Y}\right)\left(  \boldsymbol{\Delta}_{1} + (1-\gamma) \boldsymbol{K}_{Y}  \right)^{-1} \nonumber \\
& \quad + \gamma \left(  \boldsymbol{\Delta}_{1} + (1-\gamma) \boldsymbol{K}_{Y}  \right)^{-1} -\boldsymbol{\Delta}_{1}^{-1}\\
& = (1-\gamma)\left(  \boldsymbol{\Delta}^{-1}_{1} +  \boldsymbol{K}^{-1}_{Y} \right)\left(  (1-\gamma)\boldsymbol{\Delta}^{-1}_{1} +  \boldsymbol{K}^{-1}_{Y}  \right)^{-1}J(X+W | V)\left(  (1-\gamma)\boldsymbol{\Delta}^{-1}_{1} +  \boldsymbol{K}^{-1}_{Y}  \right)^{-1}\left(  \boldsymbol{\Delta}^{-1}_{1} +  \boldsymbol{K}^{-1}_{Y} \right) \nonumber \\
& \quad + \gamma \left(  \boldsymbol{\Delta}_{1} + (1-\gamma) \boldsymbol{K}_{Y}  \right)^{-1}-\boldsymbol{\Delta}_{1}^{-1}\\
&\overset{(b)}=\frac{1-\gamma}{\gamma}\left(  \boldsymbol{\Delta}^{-1}_{1} +  \boldsymbol{K}^{-1}_{Y} \right)\left(  (1-\gamma)\boldsymbol{\Delta}^{-1}_{1} +  \boldsymbol{K}^{-1}_{Y}  \right)^{-1}\left(  \boldsymbol{\Delta}^{-1}_{1} +  \boldsymbol{K}^{-1}_{Y} \right)+ \gamma \left(  \boldsymbol{\Delta}_{1} + (1-\gamma) \boldsymbol{K}_{Y}  \right)^{-1}-\boldsymbol{\Delta}_{1}^{-1} \nonumber \\
& \quad -\frac{1-\gamma}{\gamma^2}\left(  \boldsymbol{\Delta}^{-1}_{1} +  \boldsymbol{K}^{-1}_{Y} \right) \cov (X |X + W, V)\left(  \boldsymbol{\Delta}^{-1}_{1} +  \boldsymbol{K}^{-1}_{Y} \right)\\
&=\frac{1-\gamma}{\gamma} \left(  \boldsymbol{\Delta}^{-1}_{1} +  \boldsymbol{K}^{-1}_{Y}     \right)-\frac{1-\gamma}{\gamma^2}\left(  \boldsymbol{\Delta}^{-1}_{1} +  \boldsymbol{K}^{-1}_{Y} \right) \cov (X |X + W, V)\left(  \boldsymbol{\Delta}^{-1}_{1} +  \boldsymbol{K}^{-1}_{Y} \right),\label{eq:565}
\end{align}
where in (a) $W$ is a Gaussian random vector with covariance $\boldsymbol{K}_{W} = \gamma \left(       (1-\gamma)\boldsymbol{\Delta}_{1}^{-1} + \boldsymbol{K}_{Y}^{-1}     \right)^{-1}$, and (b) is due to complementary identity of Lemma \ref{lea1} in Appendix \ref{app_lea2}.

Notice that $\boldsymbol{K}_{W} \preceq \boldsymbol{K}_{Y}$, then
applying Lemma \ref{MMSE_inq} in Appendix \ref{app_lea2}, we have lower bounds on $\cov (X | X+W, V)$,
\begin{align}
&\cov(X|X+W,V)^{-1}\nonumber \\
 & \succeq \cov(X|X+N_{Y},V)^{-1} + \boldsymbol{K}_{W}^{-1} - \boldsymbol{K}_{Y}^{-1} \\
&= \cov(X|X+N_{Y},V)^{-1} +\frac{1-\gamma}{\gamma} \left( \boldsymbol{\Delta}_{1}^{-1} +  \boldsymbol{K}_{Y}^{-1} \right)\\
& \overset{(a)}\succeq \boldsymbol{D}^{-1} + \frac{1-\gamma}{\gamma} \left(  \boldsymbol{\Delta}^{-1}_{1} +  \boldsymbol{K}^{-1}_{Y}     \right)\label{eq:567}
\end{align}
where (a) is due to the MMSE distortion constraint $\cov (X | Y, V)\preceq \boldsymbol{D}$.

Substituting \eqref{eq:565} and \eqref{eq:567} into \label{eq:te3}, we obtain
\begin{align}
&\tr \left\{   \boldsymbol{M}_{3}\left(  J \left(  \left. X^{(1)}_{+, \gamma}  \right| X^{(1)}_{-, \gamma}+\sqrt{\gamma}N_{Y} - \sqrt{1-\gamma}N_{Y}^{G}, V  \right) - \boldsymbol{\Delta}^{-1}_{1}  \right)           \right\} \nonumber \\
& \geq \tr \left\{ \boldsymbol{M}_{3}\left(  \frac{1-\gamma}{\gamma} \left(  \boldsymbol{\Delta}^{-1}_{1} +  \boldsymbol{K}^{-1}_{Y}     \right)-\frac{1-\gamma}{\gamma^2}\left(  \boldsymbol{\Delta}^{-1}_{1} +  \boldsymbol{K}^{-1}_{Y} \right) \left( \boldsymbol{D}^{-1} + \frac{1-\gamma}{\gamma} \left(  \boldsymbol{\Delta}^{-1}_{1} +  \boldsymbol{K}^{-1}_{Y}     \right)\right)^{-1}  \left(  \boldsymbol{\Delta}^{-1}_{1} +  \boldsymbol{K}^{-1}_{Y} \right)                \right)           \right\}\\
&=\frac{1-\gamma}{\gamma} \tr \left\{  \boldsymbol{M}_{3}   \left(  \boldsymbol{\Delta}^{-1}_{1} +  \boldsymbol{K}^{-1}_{Y} \right) \left( \gamma \boldsymbol{D}^{-1} + {(1-\gamma)} \left(  \boldsymbol{\Delta}^{-1}_{1} +  \boldsymbol{K}^{-1}_{Y}     \right)\right)^{-1}  \left( \gamma \boldsymbol{D}^{-1} + {(1-\gamma)} \left(  \boldsymbol{\Delta}^{-1}_{1} +  \boldsymbol{K}^{-1}_{Y}     \right) - \left(  \boldsymbol{\Delta}^{-1}_{1} +  \boldsymbol{K}^{-1}_{Y} \right)\right)    \right\}\\
&=(1-\gamma)\tr \left\{  \boldsymbol{M}_{3}   \left(  \boldsymbol{\Delta}^{-1}_{1} +  \boldsymbol{K}^{-1}_{Y} \right) \left( \gamma \boldsymbol{D}^{-1} + {(1-\gamma)} \left(  \boldsymbol{\Delta}^{-1}_{1} +  \boldsymbol{K}^{-1}_{Y}     \right)\right)^{-1}  \left(  \boldsymbol{D}^{-1} -  \boldsymbol{\Delta}^{-1}_{1} -  \boldsymbol{K}^{-1}_{Y} \right)    \right\}\\
&\overset{(a)}=0,\label{eq:tter3}
\end{align}
where (a) is due to KKT condition $\left(  \boldsymbol{D}^{-1} -  \boldsymbol{\Delta}^{-1}_{1} -  \boldsymbol{K}^{-1}_{Y} \right)\boldsymbol{M}_{3}=0$ in \eqref{eq:KKT5}.

Then Combining \eqref{eq:tter1}, \eqref{eq:tter2} and \eqref{eq:tter3}, this completes the perturbation proof of $dg(\gamma)/d\gamma \leq 0$, and so the extremal inequality \eqref{eq:exinq} in Theorem \ref{ext_thm}.

\section{Conclusion}
Extremal inequalities has been found to be important to establish the converse result for the vector Gaussian network information theory problems. However, there are several instances so that the conventional perturbation approach in original probability space is resisted to show Gaussian optimality. In this paper, we develop a new method of constructing the monotone path in tensorized probability space. Several classical extremal inequalities are shown to be established via standard perturbation approach under this construction. As applications, the capacity region of the MIMO Gaussian broadcast channel and the rate-distortion-equivocation function of the vector Gaussian secure source coding are also revisited. It is shown that the new method monotone path construction is more flexible and powerful than original construction in single space. We expect the new extremal inequality approach to play important roles in solving other Gaussian network communication problems in the future.

%
%

\appendices

\section{Preliminaries on Fisher Information} \label{app_lea2}

We begin with the definition of conditional Fisher information matrix and MMSE matrix.
\smallskip
\begin{definition}
Let $({X}, U)$ be a pair of jointly distributed random vectors with differentiable conditional probability density function:
\begin{equation}
f(\boldsymbol{x}|u) \triangleq f(x_{1},\ldots,x_{n}|u).
\end{equation}
The vector-valued score function is defined as
\begin{equation}
\nabla \log f(\boldsymbol{x}| u)  = \left[\frac{\partial \log f(\boldsymbol{x}| u)}{\partial x_{1}}, \cdots, \frac{\partial \log f(\boldsymbol{x}| u)}{\partial x_{n}} \right]^{T}.
\end{equation}
The conditional Fisher information of $X$ respect to $U$ is given by
\begin{equation}
J(X| U) = \mathbb{E}\left[  \left(\nabla \log f(\boldsymbol{x}| u) \right) \cdot \left(\nabla \log f(\boldsymbol{x} | u) \right)^{T}    \right ].
\end{equation}
\end{definition}

\smallskip
\begin{definition}\label{def_MMSE}
Let $(X,Y,U)$ be a set of jointly distributed random vectors. The conditional covariance matrix of $X$ given $(Y,U)$ is defined as
\begin{equation}
\cov (X|Y,U) = \mathbb{E}\left[    \left( X - \mathbb{E}[X|Y,U] \right)  \cdot    \left( X - \mathbb{E}[X|Y,U] \right)^{T}           \right].
\end{equation}
\end{definition}

\smallskip
\begin{lemma} [Matrix Version of de Bruijn's Identity]\label{de}
Let $({X}, U)$ be a pair of jointly distributed random vectors, and ${N} \thicksim {N}(\mathbf{0}, \boldsymbol{\Sigma})$ be a Gaussian random vector independent of  $({X}, U)$. Then
 \begin{equation}
 \nabla_{\boldsymbol{\Sigma} }h({X}+  {N} | U) = \frac{1}{2}  J ({X}+{N} | U) . \label{eq:de}
 \end{equation}
\end{lemma}

Lemma \ref{de} is a conditional version of \cite[Theorem 1]{PV06}, which provides a link between differential entropy and Fisher information.
\smallskip
\begin{lemma} \label{fi_inq}
Let $({X}, {Y},U)$ be a set of jointly distributed random vectors. Assume that ${X}$ and ${Y}$ are
conditionally independent given $U$.  Then for any square matrix $\boldsymbol{A}$ and $\boldsymbol{B}$,
\begin{equation}
(\boldsymbol{A}+ \boldsymbol{B})J({X}+ {Y} | U)(\boldsymbol{A}+ \boldsymbol{B})^{T} \preceq \boldsymbol{A} J({X} | U)\boldsymbol{A}^{T} + \boldsymbol{B} J({Y} | U)\boldsymbol{B}^{T}. \label{inq:bod}
\end{equation}
\end{lemma}
\begin{IEEEproof}
From the conditional version of matrix Fisher information inequality in \cite[Appendix II]{LV07}, we have
\begin{equation}
J(X+Y|U) \preceq \boldsymbol{K}J(X|U)\boldsymbol{K}^{T} + (\boldsymbol{I} - \boldsymbol{K})J(Y|U)(\boldsymbol{I} - \boldsymbol{K})^{T},
\end{equation}
for any square matrix $\boldsymbol{K}$. Setting
\begin{equation}
\boldsymbol{K} = (\boldsymbol{A}+ \boldsymbol{B})^{-1}\boldsymbol{A}
\end{equation}
proves  \eqref{inq:bod}.
\end{IEEEproof}
\smallskip
\begin{lemma} \label{MMSE_inq}
Let $X$ be a Gaussian random vector and $U$ be an arbitrary random vector. Let $N_{1}$ and $N_{2}$ be two zero-mean Gaussian random vectors, independent of $(X,U)$, with covariance matrices $\boldsymbol{\Sigma}_{1}$ and $\boldsymbol{\Sigma}_{2}$, respectively. If
\begin{equation}
\boldsymbol{\Sigma}_{2} \succ \boldsymbol{\Sigma}_{1} \succ \boldsymbol{0},
\end{equation}
then
\begin{equation}
\cov \left( X \big| X+N_{1},U\right)^{-1} - \boldsymbol{\Sigma}_{1}^{-1} \succeq \cov \left( X \big| X+N_{2},U\right)^{-1} - \boldsymbol{\Sigma}_{2}^{-1}.
\end{equation}
\end{lemma}
\smallskip

\begin{lemma} [Cram\'{e}r--Rao Inequality] \label{cri}
Let $({X}, U)$ be a pair of jointly distributed random vectors. Assume that the conditional covariance matrix $ \cov (X| U) \succ \mathbf{0}$, then
$$
J({X}| U)^{-1} \preceq \cov ({X}| U).
$$
\end{lemma}

One can refer to the proof of unconditional version in \cite[Theorem 20]{DCT91}.
\smallskip
\begin{lemma} [Data Processing Inequality] \label{DP_FI}
Let $({X}, U, V)$ be a set of jointly distributed random vectors. Assume that $U \rightarrow V \rightarrow {X}$ form a Markov chain. Then
\begin{equation}
J({X} | U) \preceq J({X} | V).
\end{equation}
\end{lemma}

Lemma \ref{DP_FI} is analogous to \cite[Lemma 3]{Z98}, and can be easily proved using the chain rule of Fisher information matrix \cite[Lemma 1]{Z98}.

\begin{lemma}\label{lea1}
Let $(X, U)$ be a pair of jointly distributed random vectors, and $N \thicksim \mathcal{N}(\boldsymbol{0}, \boldsymbol{\Sigma})$ be a Gaussian random vector independent of  $(X, U)$.  Then
\begin{equation}
 J(X + N | U) + \boldsymbol{\Sigma}^{-1} \cov( X | X +  N, U) \boldsymbol{\Sigma}^{-1} = \boldsymbol{\Sigma}^{-1}. \label{eq_lea1}
 \end{equation}
\end{lemma}
\begin{remark}
The complementary identity in Lemma \ref{lea1} provides a link between Fisher information and MMSE, and its proof can be found in \cite[Corollary 1]{PV06}.
\end{remark}
\smallskip

\begin{lemma}\label{lea2}
Let $(X, U)$ be a pair of jointly distributed random vectors. Let $N_{1} \thicksim \mathcal{N}(\boldsymbol{0}, \boldsymbol{\Sigma}_{1})$ and $N_{2} \thicksim \mathcal{N}(\boldsymbol{0}, \boldsymbol{\Sigma}_{2})$ be two mutually independent Gaussian random vectors, and they are independent of $(X,U)$ as well. We have
\begin{equation}\label{eq_lea2}
J(X+N_{0}|U)+\left( \boldsymbol{\Sigma}_{1}^{-1} +  \boldsymbol{\Sigma}_{2}^{-1} \right) \cov(X|X+N_{1},X+N_{2},U) \left( \boldsymbol{\Sigma}_{1}^{-1} +  \boldsymbol{\Sigma}_{2}^{-1} \right) = \left( \boldsymbol{\Sigma}_{1}^{-1} +  \boldsymbol{\Sigma}_{2}^{-1} \right)
\end{equation}
where
\begin{equation}
N_{0}= \left( \boldsymbol{\Sigma}_{1}^{-1} + \boldsymbol{\Sigma}_{2}^{-1}\right)^{-1} \left(  \boldsymbol{\Sigma}_{1}^{-1} N_{1} +    \boldsymbol{\Sigma}_{2}^{-1} N_{2} \right),
\end{equation}
is a Gaussian random vector with zero mean and covariance $\left( \boldsymbol{\Sigma}_{1}^{-1} + \boldsymbol{\Sigma}_{2}^{-1}\right)^{-1}$.
\end{lemma}

\begin{IEEEproof}
Beginning with the expression of MMSE in Definition \ref{def_MMSE}, we denote that
\begin{align}
& \cov(X|X+N_{1},X+N_{2},U)  \nonumber \\
&\quad= \boldsymbol{\Sigma}_{2}\cov \left( \boldsymbol{\Sigma}_{2}^{-1} X \left| \boldsymbol{\Sigma}_{1}^{-1} (X+N_{1}) + \boldsymbol{\Sigma}_{2}^{-1} (X+N_{2}), X+N_{1},X+N_{2},U \right.\right)\boldsymbol{\Sigma}_{2}\\
&\quad= \boldsymbol{\Sigma}_{2} \cov \left( \boldsymbol{\Sigma}_{2}^{-1} X \left| X + \left( \boldsymbol{\Sigma}_{1}^{-1} + \boldsymbol{\Sigma}_{2}^{-1}\right)^{-1}   \boldsymbol{\Sigma}_{1}^{-1} N_{1} +  \left(  \boldsymbol{\Sigma}_{1}^{-1} + \boldsymbol{\Sigma}_{2}^{-1}\right)^{-1}   \boldsymbol{\Sigma}_{2}^{-1} N_{2} , X+N_{1},X+N_{2},U \right. \right)\boldsymbol{\Sigma}_{2}\label{eq_cov}
\end{align}

Since $N_{i}, i=1,2$, are zero mean Gaussian random vectors with positive definite covariance matrices $\boldsymbol{\Sigma}_{i}, i=1,2$, and denote that
\begin{align}
&{N}_{0} \triangleq \mathbb{E}\left[ N_{i}           \left|  \left( \boldsymbol{\Sigma}_{1}^{-1} + \boldsymbol{\Sigma}_{2}^{-1}\right)^{-1}   \boldsymbol{\Sigma}_{1}^{-1} N_{1} +  \left(  \boldsymbol{\Sigma}_{1}^{-1} + \boldsymbol{\Sigma}_{2}^{-1}\right)^{-1}   \boldsymbol{\Sigma}_{2}^{-1} N_{2} \right. \right] \nonumber \\
&\quad = \left( \boldsymbol{\Sigma}_{1}^{-1} + \boldsymbol{\Sigma}_{2}^{-1}\right)^{-1} \left(  \boldsymbol{\Sigma}_{1}^{-1} N_{1} +    \boldsymbol{\Sigma}_{2}^{-1} N_{2} \right), \qquad i=1,2.
\end{align}
It is known that $N_{i}$ can be decomposed as
\begin{equation}
N_{i}= {N}_{0} + Q_{i}, \qquad i=1,2.
\end{equation}
where $Q_{i}, i=1,2$, are mutually independent zero mean Gaussian random vectors with positive definite covariance matrices $\tilde{\boldsymbol{\Sigma}}_{i}, i=1,2$, given by
\begin{align}
\tilde{\boldsymbol{\Sigma}}_{i}=\boldsymbol{\Sigma}_{i}  \left( \boldsymbol{\Sigma}_{1} + \boldsymbol{\Sigma}_{2}\right)^{-1}\boldsymbol{\Sigma}_{i},  \qquad i=1,2,
\end{align}
and are independent of ${N}_{0}$.

It is because the following long Markov Chain:
\begin{equation}
U \rightarrow X \rightarrow X+ {N}_{0} \rightarrow (X+N_{1},X+N_{2}),
\end{equation}
the expression of MMSE in \eqref{eq_cov} can be simplified as
\begin{align}
&\cov(X|X+N_{1},X+N_{2},U)\nonumber \\
& \quad = \boldsymbol{\Sigma}_{2} \cov \left(\boldsymbol{\Sigma}_{2}^{-1}X \left| X + {N}_{0},U \right.\right) \boldsymbol{\Sigma}_{2} \\
& \quad = \boldsymbol{\Sigma}_{2} \cov \left(\boldsymbol{\Sigma}_{2}^{-1}X  + \boldsymbol{\Sigma}_{1}^{-1} (X + {N}_{0}) \left| X + {N}_{0},U \right.\right)\boldsymbol{\Sigma}_{2} \\
&\quad = \boldsymbol{\Sigma}_{2} \cov \left(\left( \boldsymbol{\Sigma}_{1}^{-1}+ \boldsymbol{\Sigma}_{2}^{-1} \right) X  + \boldsymbol{\Sigma}_{1}^{-1} {N}_{0} \left| X + {N}_{0},U \right.\right) \boldsymbol{\Sigma}_{2} \\
&\quad = \boldsymbol{\Sigma}_{2} \cov \left(\left( \boldsymbol{\Sigma}_{1}^{-1}+ \boldsymbol{\Sigma}_{2}^{-1} \right) X  + \boldsymbol{\Sigma}_{1}^{-1} {N}_{0} + \boldsymbol{\Sigma}_{1}^{-1}Q_{1} \left| X + {N}_{0},U \right.\right)\boldsymbol{\Sigma}_{2} - \boldsymbol{\Sigma}_{2} \left( \boldsymbol{\Sigma}_{1} + \boldsymbol{\Sigma}_{2}\right)^{-1}\boldsymbol{\Sigma}_{2} \\
&\quad=\boldsymbol{\Sigma}_{2}  \cov \left(\left( \boldsymbol{\Sigma}_{1}^{-1}+ \boldsymbol{\Sigma}_{2}^{-1} \right) X  + \boldsymbol{\Sigma}_{1}^{-1} {N}_{1}  \left|\left(\boldsymbol{\Sigma}_{1}^{-1}+ \boldsymbol{\Sigma}_{2}^{-1} \right) X + \boldsymbol{\Sigma}_{1}^{-1} {N}_{1}+ \boldsymbol{\Sigma}_{2}^{-1} {N}_{2},U \right.\right)\boldsymbol{\Sigma}_{2}- \boldsymbol{\Sigma}_{2}\left( \boldsymbol{\Sigma}_{1} + \boldsymbol{\Sigma}_{2}\right)^{-1}\boldsymbol{\Sigma}_{2} \label{eq:ci1}
\end{align}

Now making use of complementary identity \eqref{eq_lea1} in Lemma \ref{lea1}, we have
\begin{align}
& \boldsymbol{\Sigma}_{2}  \cov \left(\left( \boldsymbol{\Sigma}_{1}^{-1}+ \boldsymbol{\Sigma}_{2}^{-1} \right) X  + \boldsymbol{\Sigma}_{1}^{-1} {N}_{1}  \left|\left(\boldsymbol{\Sigma}_{1}^{-1}+ \boldsymbol{\Sigma}_{2}^{-1} \right) X + \boldsymbol{\Sigma}_{1}^{-1} {N}_{1}+ \boldsymbol{\Sigma}_{2}^{-1} {N}_{2},U \right.\right)\boldsymbol{\Sigma}_{2} \nonumber\\
& \quad= \boldsymbol{\Sigma}_{2} - J(\left(\boldsymbol{\Sigma}_{1}^{-1}+ \boldsymbol{\Sigma}_{2}^{-1} \right) X + \boldsymbol{\Sigma}_{1}^{-1} {N}_{1}+ \boldsymbol{\Sigma}_{2}^{-1} {N}_{2}|U) \\
& \quad= \boldsymbol{\Sigma}_{2}- \left(\boldsymbol{\Sigma}_{1}^{-1}+ \boldsymbol{\Sigma}_{2}^{-1} \right)^{-1} J (X+N_{0}|U)\left(\boldsymbol{\Sigma}_{1}^{-1}+ \boldsymbol{\Sigma}_{2}^{-1} \right)^{-1} \label{eq:ci2}
\end{align}

At last, we combine \eqref{eq:ci1} and \eqref{eq:ci2} together, and further invoke the Woodbury matrix inversion lemma on $ \boldsymbol{\Sigma}_{i}, i=1,2$, such that
\begin{equation}
 \left( \boldsymbol{\Sigma}_{1}^{-1} +  \boldsymbol{\Sigma}_{2}^{-1} \right)^{-1} =  \boldsymbol{\Sigma}_{2} - \boldsymbol{\Sigma}_{2}\left( \boldsymbol{\Sigma}_{1} + \boldsymbol{\Sigma}_{2}\right)^{-1}\boldsymbol{\Sigma}_{2}
\end{equation}

It can be verified that
\begin{equation}
J(X+N_{0}|U)+\left( \boldsymbol{\Sigma}_{1}^{-1} +  \boldsymbol{\Sigma}_{2}^{-1} \right) \cov(X|X+N_{1},X+N_{2},U) \left( \boldsymbol{\Sigma}_{1}^{-1} +  \boldsymbol{\Sigma}_{2}^{-1} \right) = \left( \boldsymbol{\Sigma}_{1}^{-1} +  \boldsymbol{\Sigma}_{2}^{-1} \right)
\end{equation}
where $N_{0}= \left( \boldsymbol{\Sigma}_{1}^{-1} + \boldsymbol{\Sigma}_{2}^{-1}\right)^{-1} \left(  \boldsymbol{\Sigma}_{1}^{-1} N_{1} +    \boldsymbol{\Sigma}_{2}^{-1} N_{2} \right)$. This completes the proof of Lemma \ref{lea2}.

\end{IEEEproof}
\smallskip

\begin{corollary}\label{comp}
Let $(X, U)$ be a pair of jointly distributed random vectors. Let $N_{1} \thicksim \mathcal{N}(\boldsymbol{0}, \boldsymbol{\Sigma}_{1})$ and $N_{2} \thicksim \mathcal{N}(\boldsymbol{0}, \boldsymbol{\Sigma}_{2})$ be two mutually independent Gaussian random vectors, and they are independent of $(X,U)$ as well. We have
\begin{equation}\label{eq_lee3}
J(X+N_{0}|U)= \boldsymbol{\Sigma}_2^{-1} \left(\boldsymbol{\Sigma}_1+\boldsymbol{\Sigma}_2 \right) J(X+N_1 | X+N_2, U) \left(\boldsymbol{\Sigma}_1+\boldsymbol{\Sigma}_2 \right)\boldsymbol{\Sigma}_2^{-1}-\boldsymbol{\Sigma}_2^{-1} \left(\boldsymbol{\Sigma}_1+\boldsymbol{\Sigma}_2 \right)\boldsymbol{\Sigma}_2^{-1},
\end{equation}
where
\begin{equation}
N_{0}= \left( \boldsymbol{\Sigma}_{1}^{-1} + \boldsymbol{\Sigma}_{2}^{-1}\right)^{-1} \left(  \boldsymbol{\Sigma}_{1}^{-1} N_{1} +    \boldsymbol{\Sigma}_{2}^{-1} N_{2} \right),
\end{equation}
is a Gaussian random vector with zero mean and covariance $\left( \boldsymbol{\Sigma}_{1}^{-1} + \boldsymbol{\Sigma}_{2}^{-1}\right)^{-1}$.
\end{corollary}
\begin{IEEEproof}
From Lemma \ref{lea2}, we write
\begin{align}
&J(X+N_{0}|U)\nonumber \\
&=\left( \boldsymbol{\Sigma}_{1}^{-1} +  \boldsymbol{\Sigma}_{2}^{-1} \right) - \left( \boldsymbol{\Sigma}_{1}^{-1} +  \boldsymbol{\Sigma}_{2}^{-1} \right) \cov(X|X+N_{1},X+N_{2},U) \left( \boldsymbol{\Sigma}_{1}^{-1} +  \boldsymbol{\Sigma}_{2}^{-1} \right)\\
&=\left( \boldsymbol{\Sigma}_{1}^{-1} +  \boldsymbol{\Sigma}_{2}^{-1} \right) -\left( \boldsymbol{\Sigma}_{1}^{-1} +  \boldsymbol{\Sigma}_{2}^{-1} \right)\left( \boldsymbol{\Sigma}_{1} -\boldsymbol{\Sigma}_{1} J(X+N_1 | X+N_{2}, U) \boldsymbol{\Sigma}_{1}\right)\left( \boldsymbol{\Sigma}_{1}^{-1} +  \boldsymbol{\Sigma}_{2}^{-1} \right)\label{eq:ttt}\\
&=\boldsymbol{\Sigma}_2^{-1} \left(\boldsymbol{\Sigma}_1+\boldsymbol{\Sigma}_2 \right) J(X+N_1 | X+N_2, U) \left(\boldsymbol{\Sigma}_1+\boldsymbol{\Sigma}_2 \right)\boldsymbol{\Sigma}_2^{-1}-\boldsymbol{\Sigma}_2^{-1} \left(\boldsymbol{\Sigma}_1+\boldsymbol{\Sigma}_2 \right)\boldsymbol{\Sigma}_2^{-1},
\end{align}
where \eqref{eq:ttt} is by applying Lemma \ref{lea1} by treating $(X+N_{2}, U)$ as $U$ in \eqref{eq_lea1}.
\end{IEEEproof}

\section{Proof of Proposition \ref{seq}}\label{proof_seq}
\subsubsection{Proof of \eqref{eq:81}}
From condition of \eqref{eq:57}, it can be easily observed that $\sum_{i=1}^{K} \mu_i \boldsymbol{A}^{(1)}_{i} = 0$. Now assume
\begin{equation}
\sum_{i=j}^{K} \mu_i \boldsymbol{A}^{(j)}_{i} = 0,
\end{equation}
for $1\leq j \leq L-1$. From \eqref{eq:80}, it can be verified that
\begin{align}
\sum_{i=j+1}^{L} \mu_i \boldsymbol{A}^{(j+1)}_{i} &= \sum_{i=j+1}^{K} \mu_{i} \left(  \boldsymbol{A}^{(j)}_{i} + \frac{\mu_{j}}{\sum_{k=j+1}^{L} \mu_{k}} \boldsymbol{A}^{(j)}_{j}\right) \\
&= \sum_{i=j+1}^{L} \mu_{i}\boldsymbol{A}^{(j)}_{i} + \mu_{j} \boldsymbol{A}^{(j)}_{j}\\
&=0.
\end{align}
This completes the proof of $\sum_{i=j}^{L} \mu_i \boldsymbol{A}^{(j)}_{i} = 0$ via induction.
\subsubsection{Proof of \eqref{eq:82}} From the construction of $\boldsymbol{A}^{(j)}_{i}$ in \eqref{eq:80} and the fact that $\boldsymbol{K}_{1} \preceq \ldots \preceq \boldsymbol{K}_{L}$, it can be shown that
\begin{equation}
\boldsymbol{A}^{(j)}_{j}\succeq\ldots \succeq \boldsymbol{A}^{(j)}_{L}, \quad j=1,\ldots,L.
\end{equation}
So, we have
\begin{align}
\sum_{i=j}^{L} \mu_i \boldsymbol{A}^{(j)}_{j} \succeq \sum_{i=j}^{L} \mu_i \boldsymbol{A}^{(j)}_{i}=0.
\end{align}
Since $\sum_{i=j}^{L} \mu_i \geq 0$, it implies that $\boldsymbol{A}^{(j)}_{j} \succeq 0$.
\smallskip



\bibliographystyle{IEEEtran}
\bibliography{EI}

\end{document}